\documentclass[aps,prl,reprint,balancelastpage,nofootinbib,preprintnumbers,superscriptaddress]{revtex4-1}

\usepackage{slashed}
\usepackage{bm}
\usepackage{latexsym,amssymb,amsmath,float,url,mathrsfs}
\usepackage{latexsym}
\usepackage{graphicx}
\usepackage{epstopdf}
\usepackage{amsmath}
\usepackage{comment}
\usepackage{subfigure}
\usepackage{natbib}
\usepackage{pifont}
\usepackage{blindtext}
\usepackage{adjustbox}
\usepackage{multirow}
\usepackage{tabularx}
\usepackage[table]{xcolor}
\usepackage{orcidlink}
\usepackage{fontawesome}

\usepackage{blkarray}
\usepackage{graphicx}
\usepackage{amsfonts}
\usepackage{soul}
\usepackage{amssymb}
\usepackage{amsmath}
\usepackage{cancel}
\usepackage{tcolorbox}
\usepackage{tcolorbox}

\usepackage{graphics,appendix,afterpage,makecell} 

\def\LA{\mathcal{A}_{[1][2]}^{[k][3]}}

\newcommand{\Jijab}[4]{ \mathscr{I}_{\if\relax\detokenize{#1}\relax[k]\else[#1]\fi \if\relax\detokenize{#2}\relax[k]\else[#2]\fi }^{\if\relax\detokenize{#3}\relax[k]\else[#3]\fi \if\relax\detokenize{#4}\relax[k]\else[#4]\fi  }}

\newcommand{\Aijab}[4]{ A_{\if\relax\detokenize{#1}\relax[k]\else[#1]\fi \if\relax\detokenize{#2}\relax[k]\else[#2]\fi }^{\if\relax\detokenize{#3}\relax[k]\else[#3]\fi \if\relax\detokenize{#4}\relax[k]\else[#4]\fi  }}

\newcommand{\Wijab}[4]{  W_{\if\relax\detokenize{#1}\relax[k]\else[#1]\fi\if\relax\detokenize{#2}\relax[k]\else[#2]\fi }^{\if\relax\detokenize{#3}\relax[k]\else[#3]\fi\if\relax\detokenize{#4}\relax[k]\else[#4]\fi  }}

\newcommand{\dijab}[4]{  \delta_{\if\relax\detokenize{#1}\relax[k]\else[#1]\fi\if\relax\detokenize{#2}\relax[k]\else[#2]\fi }^{\if\relax\detokenize{#3}\relax[k]\else[#3]\fi\if\relax\detokenize{#4}\relax[k]\else[#4]\fi  }}

\definecolor{oucrimsonred}{rgb}{0.6, 0.0, 0.0}
\definecolor{persianblue}{rgb}{0.11, 0.22, 0.73}
\definecolor{forestgreen}{rgb}{0.13,0.35,0.13}
\definecolor{lightgray}{rgb}{0.83, 0.83, 0.83}
 \hypersetup{colorlinks, citecolor=oucrimsonred, linkcolor=black, urlcolor=oucrimsonred}
\definecolor{cornellred}{rgb}{0.7, 0.11, 0.11}
\definecolor{navyblue}{rgb}{0.0, 0.0, 0.5}
\definecolor{amethyst}{rgb}{0.6, 0.4, 0.8}
\definecolor{yellow}{rgb}{1.0, 1.0, 0.0}
\definecolor{firebrick}{rgb}{0.7, 0.13, 0.13}
\definecolor{tangerineyellow}{rgb}{1.0, 0.8, 0.0}
\definecolor{deepfuchsia}{rgb}{0.76, 0.33, 0.76}
\definecolor{amber}{rgb}{1.0, 0.75, 0.0}
\definecolor{VioletRed4}{rgb}{0.55, 0.13, .32}
\definecolor{indiagreen}{rgb}{0.07, 0.53, 0.03}
\definecolor{VioletRed4}{rgb}{0.55, 0.13, .32}
\newcommand{\be}{\begin{equation}}
\newcommand{\ee}{\end{equation}}
\newcommand{\bea}{\begin{equation} \begin{aligned}}
\newcommand{\eea}{\end{aligned} \end{equation}}

\definecolor{oucrimsonred}{rgb}{0.6, 0.0, 0.0}
\newcommand\vertarrowbox[3][6ex]{%
  \begin{array}[t]{@{}c@{}} #2 \\
  \left\uparrow\vcenter{\hrule height #1}\right.\kern-\nulldelimiterspace\\
  \makebox[0pt]{\scriptsize#3}
  \end{array}%
}

\definecolor{verdechiaro}{rgb}{0.6,1,0.6}
\definecolor{giallochiaro}{rgb}{1,1,0.6}
\definecolor{bluscuro}{rgb}{0.15, 0.2, 0.9}
\definecolor{verdes}{rgb}{0.1, 0.5, 0.1}%
\definecolor{tangerineyellow}{rgb}{1.0, 0.8, 0.0}

\definecolor{americanrose}{rgb}{1.0, 0.01, 0.24}
\definecolor{cobalt}{rgb}{0.0, 0.28, 0.67}
\definecolor{brandeisblue}{rgb}{0.0, 0.44, 1.0}
\definecolor{mycolor}{rgb}{0.0, 0.0, 0.5}
\definecolor{oxfordblue}{rgb}{0.0, 0.13, 0.28}
\definecolor{azure}{rgb}{0.0, 0.5, 1.0}
\definecolor{turquoiseblue}{rgb}{0.0, 1.0, 0.94}
\newtcolorbox{mynewbox}[1]{colback=white!5!white,colframe=azure!75!black,fonttitle=\bfseries,title=#1}
\newtcolorbox{mybox}{colback=mycolor!5!white,colframe=azure!75!black}
\newtcolorbox{mynamedbox}[1]{colback=mycolor!5!white,colframe=azure!75!black,title=#1}
\definecolor{venetianred}{rgb}{0.78, 0.03, 0.08}
\newtcolorbox{mynamedbox1}[1]{colback=venetianred!5!white,colframe=venetianred!80!black,title=#1}
\newtcolorbox{mynamedbox2}[1]{colback=azure!5!white,colframe=azure!80!black,title=#1}

\definecolor{verdes}{rgb}{0.1, 0.5, 0.1}%
\definecolor{cornellred}{rgb}{0.7, 0.11, 0.11}

\definecolor{VioletRed4}{rgb}{0.55, 0.13, .32}

\hypersetup{
     colorlinks   = true,
     citecolor    = violet,
     urlcolor     = violet,
     linkcolor    = violet}

\definecolor{rossocorsa}{rgb}{0.83, 0.0, 0.0}
\definecolor{green}{rgb}{0.0, 0.5, 0.0}
\definecolor{orange}{rgb}{0.90, 0.5, 0.0}

\def\lsim{\mathrel{\rlap{\lower4pt\hbox{\hskip0.5pt$\sim$}}
    \raise1pt\hbox{$<$}}}         
\def\gsim{\mathrel{\rlap{\lower4pt\hbox{\hskip0.5pt$\sim$}}
    \raise1pt\hbox{$>$}}}         

\usepackage[normalem]{ulem}

\usepackage{cleveref}
\crefname{equation}{Eq.}{Eqs.}
\crefname{section}{Sec.}{Secs.}
\crefname{table}{Tab.}{Tabs.}
\crefname{figure}{Fig.}{Figs.}

\usepackage{hyperref}

\begin{document}

\title[]{Nonlinear Gravity and Multipole Turbulence }

\author{A. Ianniccari\orcidlink{0009-0008-9885-7737}}
\affiliation{Department of Theoretical Physics and Gravitational Wave Science Center,  \\
24 quai E. Ansermet, CH-1211 Geneva 4, Switzerland}

\author{A. Kehagias\orcidlink{0000-0001-6080-6215
}}
\affiliation{Physics Division, National Technical University of Athens, Athens, 15780, Greece}

\author{L. Lo Bianco\orcidlink{0009-0001-8867-0062}}
\affiliation{Dipartimento di Fisica, Università degli Studi di Torino, via P. Giuria, 1 10125 Torino, Italy}

\author{A. Riotto\orcidlink{0000-0001-6948-0856}}
\affiliation{Department of Theoretical Physics and Gravitational Wave Science Center,  \\
24 quai E. Ansermet, CH-1211 Geneva 4, Switzerland}


\begin{abstract}
\noindent
We derive a kinetic Boltzmann equation characterizing the long-term statistical behavior of the turbulent dynamics of nonlinear interacting gravitational wave multipoles in Minkowski spacetime and  show that, injecting a large number of gravitons with large  multipoles  drives the system toward an inverse multipole  cascade at large times.

\end{abstract}

\maketitle

\vskip 0.3cm\noindent\textbf{{\em Introduction}} -- Quasi-normal modes (QNMs) are a hallmark of black hole (BH) physics, characterizing how these objects react to external perturbations (see Ref.~\cite{Berti:2025hly} for a recent overview). Studying this response provides access to key information about the post-merger phase of compact binaries, the ringdown, during which a newly formed BH settles into equilibrium.

While the standard description has been remarkably effective when compared with current observations, a series of recent studies~\cite{London:2014cma,Mitman:2022qdl,Cheung:2022rbm,Ma:2022wpv,Redondo-Yuste:2023seq,Cheung:2023vki,Zhu:2024rej} has highlighted that nonlinear gravitational dynamics can modify the anticipated QNM spectrum. Notably, such nonlinearities may generate extra peaks in the frequency domain, typically located at combinations of linear-mode frequencies and, in many cases, carrying substantial power.
Both analytical and numerical efforts have been dedicated to probing these nonlinear oscillations~\cite{Kehagias:2023ctr,Redondo-Yuste:2023seq,Cheung:2023vki,Perrone:2023jzq,Zhu:2024rej,Ma:2024qcv,Bourg:2024jme,Bucciotti:2024zyp,Khera:2024yrk,Kehagias:2024sgh,Bucciotti:2025rxa,bourg2025quadraticquasinormalmodesnull,BenAchour:2024skv,Kehagias:2025xzm,Ling:2025wfv,Kehagias:2025ntm,Kehagias:2025tqi,Perrone:2025zhy,Fransen:2025cgv,Kehagias:2025gvk,Ianniccari:2025avm,Singh:2025xzd}. Much of this work concentrates on the so-called quadratic QNMs with multipole $2\ell$, which arises when two modes of multipole  $\ell$ annihilate. For $\ell = 2$, these nonlinear signals may even be detectable using next-generation gravitational-wave observatories~\cite{Lagos:2024ekd,Yi_2024,PhysRevD.109.064075,shi2024detectabilityresolvabilityquasinormalmodes}.

A natural question is whether the reverse mechanism, where a high  $\ell$-mode (not necessarily a QNM) produces lower  multipole states also occurs. Since this interaction is at least of the same cubic  order, one would expect it to be present as well. Such a process would correspond to an inverse transfer of energy, ultimately feeding power into the dominant and most easily observed $\ell=2$ channel.

Indeed, numerical studies have recently indicated that nonlinear mode couplings can drive inverse cascades through resonant instabilities that channel energy from higher to lower frequencies \cite{t8}. The underlying phenomenon was first identified analytically in the context of rapidly rotating BHs, where a parametric instability can trigger turbulent gravitational dynamics \cite{Yang:2014tla}. This realization has inspired a growing body of research on turbulence-like behavior in gravitational-wave systems \cite{Galtier:2017mve,Galtier:2018vbq,Benomio:2024lev,Iuliano:2024ogr,Figueras:2023ihz,Krynicki:2025fzi,Siemonsen:2025fne,t2,t3,t5,t11,t12,Krynicki:2025fzi,t14,Kehagias:2025zws}.

A full analytic treatment of nonlinear gravitational modes remains technically demanding. Nonetheless, qualitative insights and approximate results can be obtained in the large–angular-momentum regime, commonly referred to as the eikonal limit. This limit has played a central role in the study of linear perturbations~\cite{PhysRevD.30.295,PhysRev.166.1263,Cardoso_2009,Dolan_2010,Dolan_2018,Hadar:2022xag,Fransen:2023eqj} and continues to offer a valuable window into more complicated nonlinear phenomena.

In this letter we take a step further towards the understanding of how energies and angular momenta are distributed among interacting GWs by exploiting the eikonal limit. We derive 
a kinetic Boltzmann equation that describes the long-time statistical dynamics of the turbulent behavior of the multipoles of the nonlinear gravitational interacting waves, showing that lower multipoles may be  preferred at large times, thus also explaining analytically what has been recently obtained by numerical general relativity experiments. Our calculations are done in Minkowski spacetime and therefore they are relevant  on asymptotically flat spacetimes, e.g. far enough from the BH horizon. In the main text we collect the  most relevant passages, leaving the (many) technical details to the Supplemental Material (SM).


\vskip 0.3cm\noindent\textbf{{\em The Boltzmann equation in  multipole space}} -- In this section, we derive the Boltzmann equation for gravitational waves, expanded in a harmonic basis to encompass both wavenumber
$k$ and multipole $\ell$ degrees of freedom. 
For clarity and simplicity, we restrict our 
analysis to a four-point coupling and a 
$2\to 2$ scattering interaction between gravitational wave modes. Our main objective is to identify and describe the resonant part of the nonlinear dynamics, namely the subset of mode couplings that satisfy both energy and momentum conservation. These resonant interactions are the only ones that lead to a long-term energy transfer between modes, while non-resonant terms correspond to rapidly oscillating contributions that average out over long timescales. Focusing on the resonant sector thus isolates the physically relevant processes governing the statistical evolution of the gravitational-wave spectrum. The generalization to include higher-order couplings and non-resonant corrections will be presented later on.

We begin with the Hamiltonian formalism for the four-point vertex
\begin{eqnarray}
    H &=& H_2 + H_4 = \sum_{\mathbf{k}}\omega_{\mathbf{k}}a_{\mathbf{k}}a^{*}_{\mathbf{k}} \nonumber \\
    &+& \frac{1}{2}\sum_{\mathbf{k_1},\mathbf{k_2},\mathbf{k_3}, \mathbf{k_4}}W^{\mathbf{k_1},\mathbf{k_2}}_{\mathbf{k_3}, \mathbf{k_4}}\delta^{\mathbf{k_1},\mathbf{k_2}}_{\mathbf{k_3},\mathbf{k_4}} a_{\mathbf{k_1}}a_{\mathbf{k_2}}a^{*}_{\mathbf{k_3}}a^{*}_{\mathbf{k_4}},
    \label{eq:h24}
\end{eqnarray}
where \( \omega_{\mathbf{k}} \) is the mode frequency, \( a_{\mathbf{k}} \) and \( a_{\mathbf{k}}^{*} \) are complex canonical variables to the Hamiltonian $H$, \( W^{\mathbf{k_1},\mathbf{k_2}}_{\mathbf{k_3},\mathbf{k_4}} \) is the scattering amplitude, and \( \delta^{\mathbf{k_1},\mathbf{k_2}}_{\mathbf{k_3},\mathbf{k_4}} = \delta(\mathbf{k_4}+\mathbf{k_3}-\mathbf{k_1}-\mathbf{k_2}) \) enforces momentum conservation.

We would like to notice that because the gravitational-wave dispersion relation is linear in flat Minkowski spacetime and the Einstein equations possess a special null structure, all three-wave interactions are non-resonant. Equivalently, the three-point graviton amplitude in flat Minkowski spacetime vanishes on-shell (for real $\mathbf{k}$), reflecting the absence of any physical three-wave process. Cubic terms  generate only rapidly oscillating phase shifts, which can be removed by a canonical transformation\footnote{However, this is not the case in curved backgrounds where curvature introduces effective mode coupling and allows three-wave mixing between gravitational waves \cite{t8}.}. As a result, the leading-order resonant dynamics of weakly nonlinear gravitational waves are governed entirely by four-wave interactions, described by the quartic Hamiltonian $H_4$ in Eq. (\ref{eq:h24}).  Moreover, in the latter, we kept only terms of the form 
$a_{\mathbf{k_1}} a_{\mathbf{k_2}} a_{\mathbf{k_3}}^* a_{\mathbf{k_4}}^*$ and their complex conjugates, 
since these are the only combinations that satisfy the resonance conditions 
$\omega_{\mathbf{k_1}} + \omega_{\mathbf{k_2}} = \omega_{\mathbf{k_3}} + \omega_{\mathbf{k_4}}$ and 
$\mathbf{k}_1 + \mathbf{k}_2 = \mathbf{k}_3 + \mathbf{k}_4$. 
All other quartic terms arising from the product of plane waves as in Eq. (\ref{gravv}), such as 
$a_{\mathbf{k_1}} a_{\mathbf{k_2}} a_{\mathbf{k_3}} a_{\mathbf{k_4}}$ or 
$a_{\mathbf{k_1}} a_{\mathbf{k_2}} a_{\mathbf{k_3}} a_{\mathbf{k_4}}^*$, are non-resonant as 
their phases $e^{i\Delta \omega t}$ oscillate rapidly in time and therefore average out 
in the weakly nonlinear limit. 
These non-resonant contributions can be removed by a canonical transformation, 
leaving only the resonant  terms in the interaction Hamiltonian in Eq. (\ref{eq:h24}) responsible for the  two-to-two  energy exchange among gravitational-wave modes.

To proceed, we recall that  the equation of motion follows from the Hamilton equation,
\begin{equation}
    i\dot{a}_{\mathbf{k}} = \frac{\delta H}{\delta a^{*}_{\mathbf{k}}}.
\end{equation}
In order to remove linear oscillations, we define the modes \( b_{\mathbf{k}} =( e^{i\omega_{\mathbf{k}}t + i \int\omega_{\rm{S}, \mathbf{k}}dt}\,a_{\mathbf{k}})/{\epsilon} \), where \( \epsilon \) is the perturbative parameter we will use to solve the equation of motion and $\omega_{\rm{S}}$ is defined in Eq. (\ref{omegas}). Expanding \( b_{\mathbf{k}} \) in spherical harmonics,
\be
b_{k,\ell m}= \int d\Omega_k \, b_{\mathbf{k}} \, Y^*_{\ell m }(\Omega_k),
\ee
yields the evolution equation
\begin{eqnarray}
     i \dot{b}_{k,\ell m} &=& \epsilon^2 \sum_{\substack{\ell_i,m_i \\ i=1,2,3}} \prod_{i=1}^3 \int dk_i k_i^2 \, 
     \LA f(k_1,k_2,k_3) \nonumber \\
     &\cdot& b_{1,\ell_1 m_1}b_{2,\ell_2m_2}b_{3,\ell_3m_3}^{*} \, e^{i\int\omega^{\mathbf{3},\mathbf{k}}_{\mathbf{1},\mathbf{2}} dt}.
\end{eqnarray}
The coefficients $\omega^{\mathbf{3},\mathbf{k}}_{\mathbf{1},\mathbf{2}}$  , $f(k_1,k_2,k_3)$ and \( \LA \) are defined in Eqs. (\ref{omega123}), (\ref{fdiff}) and (\ref{Aterm}) respectively.

We can  expand \( b_{k,\ell m} \) around an intermediate time \( T \)  as 
\begin{equation}
    b_{k,\ell m}(T) = b_{k,\ell m }^{(0)} + \epsilon^2 b_{k,\ell m }^{(1)} + \epsilon^4 b_{k,\ell m }^{(2)} + {\cal O}(\epsilon^6),
\end{equation}
with \( b_{k,\ell m }^{(0)} \) a constant background. The intermediate time \( T \), is chosen such that \( \tau_{\rm{L}} \ll T \ll \tau_{\rm{NL}} \) where  $\tau_{\rm L}=2\pi/\omega_\mathbf{k}$  is the linear timescale associated with the characteristic scale of the linear evolution, and   $\tau_{\rm NL}$ is the nonlinear timescale,  which is the characteristic time for weak nonlinear interactions to cause an appreciable change in the wave amplitude or phase. Since the time evolution of $b_{\mathbf{k}}^{(1)}$ is proportional to $\epsilon^2$, the nonlinear timescale refers to the leading nonlinear correction  $E_{\rm NL}$ of the energy density  which is of order $\epsilon^4$ since $E_{\rm NL}\sim \omega_{\mathbf{k}}(b_\mathbf{k}^{*(0)}b_\mathbf{k}^{(1)}+c.c)\sim \epsilon^4\omega_{\mathbf{k}}$.   Therefore, the nonlinear  timescale turns out to be $\tau_{\rm NL}=2\pi/(\epsilon^4 \omega_k)$. 
In the SM, we will estimate such times for the case of the QNMs generated by a BH of mass $M$, $\tau_{\rm{L}}\sim (M/\ell )\ll \tau_{\rm{NL}}\sim M$ for $\ell\gg 1$, that is for the eikonal limit we will consider later on. In such a case $\epsilon\sim \ell^{-1/4}$.

Following Ref. \cite{t12}, we treat \( b_{k,\ell m} \) as a stochastic variable and write it as \( b_{k_i,\ell_i m_i} = \sqrt{J_{k_i,\ell_i m_i}} \, e^{i\phi_i} \), where \( J_{k_i,\ell_im_i} \in \mathbb{R}^+ \) is the intensity and \( \phi_i \) is the phase. We assume no correlation for the intensity and phase of each mode and for the  phases of the different modes.

After performing statistical averages and contractions, and taking the limit \( \epsilon \to 0 \), we arrive at the Boltzmann equation in multipole space
\begin{eqnarray}
    \dot{n}_{k,\ell m} &=& 4\pi\epsilon^4 \sum_{\substack{\ell_i,m_i \\ i=1,2,3}} \prod_{i=1}^3 \int dk_i k_i^2 \, \left| \LA \right|^2 \delta(\omega^{3k}_{12}) \nonumber \\
    &\cdot& f(k_1,k_2,k_3) \, n_{k,\ell m}n_{1,\ell_1 m_1}n_{2,\ell_2 m_2}n_{3,\ell_3 m_3} \nonumber \\
    &\cdot& \left[ \frac{1}{n_{k,\ell m}} + \frac{1}{n_{3,\ell_3 m_3}} - \frac{1}{n_{1,\ell_1 m_1}} - \frac{1}{n_{2,\ell_2 m_2}} \right],\nonumber\\
&&
\end{eqnarray}
where  the wave-energy spectrum, or  the classical ``occupation number" of the mode $(k,\ell\, m)$
is defined as
\be
n_{k,\ell m } = \langle |b_{k,\ell m}|^2 \rangle.
\ee
Clearly, $n_{k,\ell m }$ is dimensionful. It is related to the quantum-mechanical 
 occupation number $n_{k,\ell m }^{(\rm q)}$, which is dimensionless and counts the number of quanta (gravitons) in a given mode, by $n_{k,\ell m }=\hbar n_{k,\ell m }^{(\rm q)}$ in the classical limit $n_{k,\ell m }^{(\rm q)}\gg1$ and $\hbar\to 0$. 

\noindent\textbf{{\em The Eikonal limit}} -- In the eikonal limit, which corresponds to small-angle scattering dominated by the exchange of large angular momentum $\ell\gg 1$, the Boltzmann equation can be symmetrized. The specific operations leading to this limit are detailed in the SM. The resulting symmetrized equation is
\begin{align}
\dot{n}_{k,\ell m} =& \frac{1}{4} \!\!\!\sum_{\substack{\ell_i,m_i \\ i=1,2,3}} \prod_{i=1}^3 \int dk_i k_i^2 \, n_{k,\ell m}n_{1,\ell_1 m_1}n_{2,\ell_2 m_2}n_{3,\ell_3 m_3} \nonumber \\
&\cdot\Bigg[ 
\Jijab{1}{2}{}{3}+\Jijab{2}{3}{3}{}-\Jijab{}{2}{1}{3}-\Jijab{1}{}{2}{3}
\Bigg],
\end{align}
where \( 
\Jijab{1}{2}{}{3}
\) is given in Eq. (\ref{Iterm}). Applying the Zakharov transformations  \cite{Zakharov}, of which an example can be found in Eq. (\ref{coordtransf}), 
to map all integrals onto the domain of the first, we seek a steady-state power-law solution in both momentum \( k \) and multipole \( \ell \)  
\begin{eqnarray}
    n_{k,\ell m} = A k^{\nu} \ell^{\mu}.
\end{eqnarray}
We consider a general dispersion relation \( \omega \sim k^{\alpha} \), noting that for gravitational waves in flat spacetime \( \alpha = 1 \). The kinetic equation in the eikonal limit then becomes
\begin{eqnarray}\label{kinltext}
\dot{n}_{k,\ell m} &=& \frac{1}{4}\prod_{i=1}^3 \int d\ell_i  \int dk_i k_i^2 \, 
\Jijab{1}{2}{}{3}\nonumber \\
&\cdot& n_{k,\ell m}n_{1,\ell_1 m_1}n_{2,\ell_2 m_2}n_{3,\ell_3 m_3} \nonumber \\
&\cdot& \Bigg[ 1 + \left( \frac{k_3}{k} \right)^x \left( \frac{\ell_3}{\ell} \right)^y - \left( \frac{k_1}{k} \right)^x \left( \frac{\ell_1}{\ell} \right)^y \nonumber\\
&-& \left( \frac{k_2}{k} \right)^x \left( \frac{\ell_2}{\ell} \right)^y \Bigg],
\end{eqnarray}
with
\begin{eqnarray}
    x &=& \alpha - 3\nu - 2\beta - 9, \nonumber \\
    y &=& -3\mu - 4.
\end{eqnarray}
This is our central result. We recover the exponent \( x \) from Ref. \cite{t12}, augmented by a power-law dependence on the multipoles. Remarkably, the multipole turbulence cascade power-law \( y \) is independent of both the scattering details and the dispersion relation, as it depends on neither \( \beta \) nor \( \alpha \).

\vskip 0.3cm\noindent\textbf{{\em Stationary solutions}} -- 
We now examine stationary solutions. First, setting \( y = 0 \) yields \( x = \alpha \) or \( x = 0 \). The solution \( x = \alpha \) corresponds to energy conservation in flat spacetime, leading to the spectral index \( \nu_E = -2\beta/3 - 3 \). This defines a positive constant energy flux, indicative of a direct energy cascade in momentum space. The solution \( x = 0 \) gives the spectral index \( \nu_N = -2\beta/3 - 3 + \alpha/3 \), which defines a constant negative particle number flux, implying an inverse energy cascade, also called inverse waveaction cascade.

Alternatively, fixing \( x = 0 \) allows us to find stationary solutions in multipole space. Choosing \( y = 1 \) ensures angular momentum conservation via the Clebsch-Gordan coefficients in the eikonal limit, yielding the spectral index \( \mu_L = -5/3 \). The corresponding flux for the total angular momentum,
\be
\Phi_\ell = -\int^{L_{\rm max}} d\ell \, \ell \, \dot n_{k ,\ell m},
\ee
is a non-zero constant only for \( \mu_L \). The second stationary solution, \( y = 0 \), gives \( \mu_N = -4/3 \). The waveaction flux in multipole space,
\be
\mathcal{N}_\ell = -\int^{L_{\rm max}} d\ell \, \dot n_{k ,\ell m},
\ee
is a non-zero constant only for \( \mu_N \).

The sign of these constants can be understood via the so-called Fjørtoft argument \cite{fjortoft}, illustrated in Fig. \ref{fig:Fluxes}, and based on conservation theorems. The spectral indices obey the hierarchy \( \mu_{TN} > \mu_{TL} > \mu_{N} > \mu_{L} \), where \( \mu_{TN} = 0 \) and \( \mu_{TL} = -1 \) are the Rayleigh-Jeans indices that preserve the occupation number and the angular momentum with respect to $\ell$. For large negative spectral indices, both fluxes are positive. Consequently, the angular momentum flux is positive for \( \mu_L \), indicating a direct cascade from small to large \( \ell \). Conversely, the waveaction flux is negative for \( \mu_N \), defining an inverse cascade from large to small \( \ell \).

\begin{figure}[h]
	\begin{center}
\includegraphics[width=.45\textwidth]{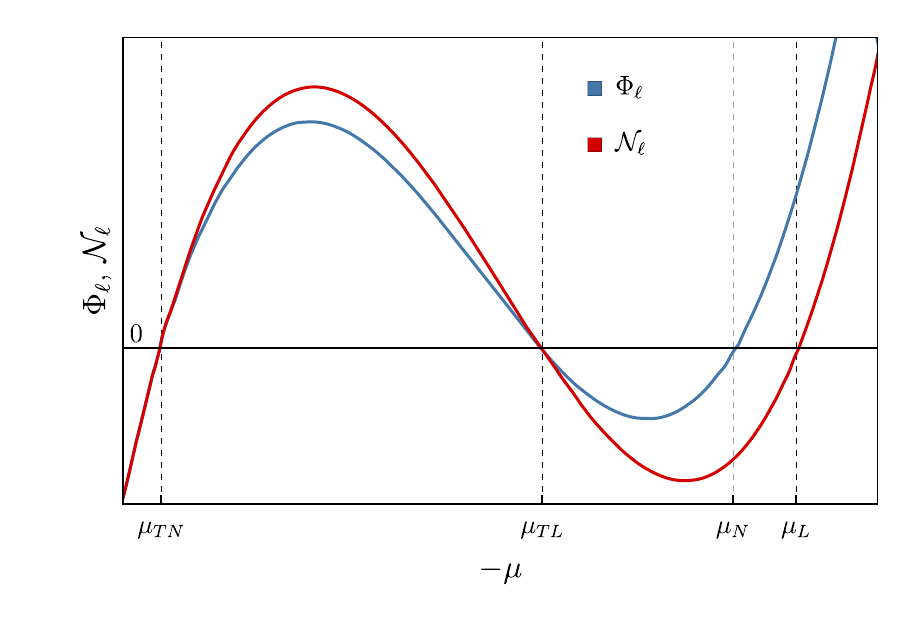}
		\caption{The Fjørtoft argument with   $\mu_{TN}=0$, $\mu_{TL}=-1$, $\mu_N=-4/3$ and $\mu_L=-5/3$. For high negative values of the spectral index both fluxes should be positives. Then the total angular momentum flux will be  positive constant in $\mu_L=-5/3$, while the flux for the number of graviton will be zero in that point, and so a constant negative in $\mu_N=-4/3$. This argument specifies the directions of both the fluxes: the flux for the number of graviton is negative, therefore the system is moving the number of particles from high  $\ell$ to smaller ones. At the contrary, the flux over the angular momentum is positive, so the system will bring the total angular momentum from small values of $\ell$ to higher values.}
        \label{fig:Fluxes} 
	\end{center}
\end{figure}
\noindent\textbf{{\em The Direction of turbulence: a matter of conservation}} --
In weak wave turbulence, the Boltzmann equation governs the slow evolution 
of the occupation number $n_{k,\ell m}$ 
due to nonlinear mode coupling. 
For a general $N$-wave interaction, this equation schematically reads
\begin{align}
\dot{n}_{k,\ell m} \sim & 
\int |T_{1,2,\ldots,N}|^2 
\Big(n_1 n_2 \ldots n_{N/2} \pm n_{N/2+1}\ldots n_N\Big)\nonumber 
\\
\cdot \ &
\delta(\omega_1+\cdots-\omega_N)
\delta(\mathbf{k}_1+\cdots-\mathbf{k}_N)\,d\mathbf{k}_1\cdots d\mathbf{k}_N,
\end{align}
where $T_{1,2,\ldots,N}$ is the vertex amplitude and the alternating $\pm$ signs 
encode the creation and destruction of particle states in the interaction. 
The parameter $N$ may be even or odd and represents the number of asymptotic 
particle states involved in the scattering vertex. 
To illustrate the scaling behavior of the collision integral, we can write symbolically
\begin{eqnarray}
\dot{n}_{k,\ell m} &\propto& \Bigg[1 \pm \left(\frac{k_{N-1}}{k}\right)^x\left(\frac{\ell_{N-1}}{\ell}\right)^y \nonumber\\
&\pm& \left(\frac{k_{N-2}}{k}\right)^x\left(\frac{\ell_{N-2}}{\ell}\right)^y \pm \ldots \Bigg],
\end{eqnarray}
where the exponents $(x,y)$ depend on the $N$-vertex theory.
This schematic form captures how the scaling of $n_{k,\ell}$ enters the 
kinetic equation and determines whether the collision integral admits 
a stationary solution. 
For even $N$, as in four-wave ($2\!\leftrightarrow\!2$) scattering, the number 
of particles is conserved, and a stationary solution exists for $(x,y)=(0,0)$, 
allowing both direct and inverse cascades. 
For odd $N$, as in three-wave ($1\!\leftrightarrow\!2$) processes, the particle number 
is not conserved, and a stationary $(x,y)=(0,0)$ solution does not exist. 
Hence, the presence or absence of conserved quantities determines the 
possible directions of turbulent transfer since direct cascades correspond to 
the flux of conserved energy toward higher $(k,\ell)$, while inverse cascades 
represent the flux of conserved waveaction toward lower $(k,\ell)$.  

The general considerations above determine whether stationary spectra
exist for different interaction vertices. We now specialize to the case of 
gravitational-wave turbulence in flat spacetime, where both energy and 
angular momentum are conserved.
In flat spacetime, where energy and angular momentum are conserved, the conditions \( (x = \alpha,\; y = 0) \) and \( (x = 0,\; y = 1) \) each yield a stationary solution for the occupation number. These conditions imply the existence of the spectral indices \( \nu_E \) and \( \mu_L \), which guarantee a direct cascade of energy and total angular momentum, respectively. Thus, the conservation of a quantity necessitates a direct cascade of its corresponding flux.

However, if the number of legs in the vertex is odd, the condition \( (x = 0,\; y = 0) \) does not provide a stationary solution. For even \( N \), a stationary solution under this condition is only possible if the number of particles created equals the number destroyed. This implies that the stationary solution for \( (x = 0,\; y = 0) \), which defines the existence of an inverse waveaction cascade via the spectral indices \( \nu_N \) and \( \mu_N \), is guaranteed solely when the scattering process conserves the number of particles.


These insights may help to elucidate the numerical results of Ref. \cite{t8}, where a
 signal has been injected toward the horizon of a spinless BH and detected the backreaction signal far from the horizon. It was observed 
an inverse cascade  in  multipole space by injecting an $\ell=6$ mode, and in energy by injecting an $\ell=2$ mode.
 In curved spacetime, angular momentum is not conserved, leaving the inverse cascade as the only viable evolutionary path. The injection of a mode with \( \ell = 6 \) corresponds to forcing the system with a high number of gravitons at large \( \ell \), thereby driving it along the inverse waveaction cascade and transferring gravitons from larger to smaller \( \ell \) values. Using the \( \ell = 2 \) mode as a driver with frequency \( \omega \) precludes a direct cascade, as energy is not conserved in the curved background. Instead, these linear modes scatter via a quartic vertex, generating nonlinear laminar modes at frequencies \( 3\omega \) and \( \omega \). The high-frequency mode \( 3\omega \) then acts as a forcing term, initiating an inverse waveaction cascade that transfers energy (manifest as the envelope amplitude) from the initial frequency \( 3\omega \) to lower frequencies over time. Crucially, the presence of an inverse cascade in both scenarios is contingent upon the nonlinear interaction conserving the number of asymptotic particle states.

\vskip 0.3cm
\noindent \textbf{{\em Conclusions}} --
Nonlinear gravitational interactions give rise to a variety of surprising phenomena, making the detection of BH QNMs a distinctive opportunity to gain insights into the fundamental nature of gravity.  Our results, along with recent numerical findings in the literature \cite{t8}, suggest that resonant interactions tend to favor the preservation of the lowest angular and frequency modes. This has important observational consequences, especially regarding the prospect of detecting quadratic QNMs.  Our findings should be extended in several directions. First of all, and maybe most importantly, the calculation should be extended close to the BH, around the photon ring. One way to facilitate the approach might be to take the Penrose limit connecting a simpler plane wave geometry to the BH photon ring, where the QNMs are located in the eikonal limit. Second of all, it would be interesting to analyze how the inverse cascades alter the power spectra of the stochastic GWs generated during the early universe along the lines of Ref. \cite{Galtier:2017mve}.

\begin{acknowledgments}
\vspace{5pt}\noindent\emph{Acknowledgments}
\vskip 0.1cm
  \noindent
  We thank F. Bernardo, D. Perrone and N. Muttoni for many useful discussions.
  A.I. and A.R.  acknowledge support from the  Swiss National Science Foundation (project number CRSII5\_213497).
\end{acknowledgments}

\bibliography{draft}

\clearpage
\newpage
\maketitle
\onecolumngrid
\begin{center}
\vspace{0.05in}

\vspace{0.1in}
{ \Large\it Supplemental Material}
\end{center}

\setcounter{equation}{0}
\setcounter{figure}{0}
\setcounter{section}{0}
\setcounter{table}{0}
\setcounter{page}{1}
\makeatletter
\renewcommand{\theequation}{S\arabic{equation}}
\renewcommand{\thefigure}{S\arabic{figure}}
\renewcommand{\thetable}{S\arabic{table}}

\section*{The Boltzmann equation for multipoles: the Four-point vertex calculation}
\noindent
Our starting point is the definition of the gravitational plane wave in Minkowski spacetime
\begin{equation}
\phantomsection
\label{gravv}
    h_{\mu\nu}(t,\vec{r})=\int \frac{d^3k}{(2\pi)^{3/2}}\sum_{s}\left[\frac{e^{-i\vec{k}\cdot\vec{r}+i\omega t}}{\sqrt{2\omega}}\epsilon^s_{\mu\nu}b^s_{\mathbf{k}}+{\rm h.c.}\right],
\end{equation}
where $s=(+,\times)$ indicates the two graviton helicities with polarization vector $\epsilon^s_{\mu\nu}$. Using the spherical harmonic expansion of the plane waves
\begin{equation}
    e^{-i\vec{k}\cdot\vec{r}}=4\pi\sum_{\ell m}(-i)^{\ell}j_\ell(kr)Y_{\ell m}(\Omega_r)Y^*_{\ell m}(\Omega_k),
\end{equation}
we can write
\begin{equation}
     h_{\mu\nu}(t,\vec{r})=4\pi \int \frac{dk\;k^2}{(2\pi)^{3/2}\sqrt{2\omega}}\sum_s\sum_{\ell m}\Big[(-i)^{\ell}j_\ell(kr)Y_{\ell m}(\Omega_r)b^s_{k,\ell m}\epsilon_{\mu\nu}^se^{+i\omega t}+{\rm h.c.}\Big],
\end{equation}
where we have defined the modes
\begin{equation}
    b_{k,\ell m}=\int d\Omega_kY^*_{\ell m}(\Omega_k)b_{\mathbf{k}}.
\end{equation}
The next step is to solve the kinetic equation for the new modes written in the multipole space. The corresponding solution will provide the whole dynamics of the graviton in coordinate space, as written in Eq.(\ref{gravv}).\\
In this section, we aim to compute the Boltzmann equation for a  \(2 \to 2\) scattering process, in which both direct and inverse cascades can be observed in the momentum modes \(k\) as well as in the multipole modes \(\ell\). To achieve this result, we first express the kinetic 
equation expanded in the harmonic basis. We begin by writing the equation of motion in the Hamiltonian formalism  for a four-point interaction vertex, for simplicity. The Hamiltonian 
corresponding to the quartic interaction can be written as follows
\begin{equation}
\phantomsection
\label{H4}
    H = H_2 +H_4 = \sum_{\mathbf{k}}\omega_{\mathbf{k}}a_{\mathbf{k}}a^{*}_{\mathbf{k}} + \frac{1}{2}\sum_{\mathbf{k_1},\mathbf{k_2},\mathbf{k_3, \mathbf{k}_4}}W^{\mathbf{k_1},\mathbf{k_2}}_{\mathbf{k_3}, \mathbf{k_4}}\delta^{\mathbf{k_1},\mathbf{k_2}}_{\mathbf{k_3},\mathbf{k_4}} a_{\mathbf{k_1}}a_{\mathbf{k_2}}a^{*}_{\mathbf{k_3}}a^{*}_{\mathbf{k_4}}\,,
\end{equation}
where $\omega_{\mathbf{k}}$ is the frequency of the mode $\mathbf{k}$, $a_{\mathbf{k}}$ and $a_{\mathbf{k}}^{*}$ are complex canonical variables to the Hamiltonian $H$, $W^{\mathbf{k_1},\mathbf{k_2}}_{\mathbf{k_3},\mathbf{k_4}}$ is the scattering amplitude and $\delta^{\mathbf{k_1},\mathbf{k_2}}_{\mathbf{k_3},\mathbf{k_4}}=\delta(\mathbf{k_4}+\mathbf{k_3}-\mathbf{k_1}-\mathbf{k_2})$ is the Dirac delta for momentum conservation.\\
The Hamilton equation is 
\begin{equation}
    i\dot{a}_{\mathbf{k}} = \frac{\delta H}{\delta a^{*}_{\mathbf{k}}}\,.
\end{equation}
Therefore, the following expression is obtained
\begin{equation}
    i \dot{a}_{\mathbf{k}} = \omega_{\mathbf{k}}a_{\mathbf{k}} + \sum_{\mathbf{k_1},\mathbf{k_2},\mathbf{k_3}}W^{\mathbf{k_1},\mathbf{k_2}}_{\mathbf{k_3},\mathbf{k}}\delta^{\mathbf{k_1},\mathbf{k_2}}_{\mathbf{k_3},\mathbf{k}} a_{\mathbf{k_1}}a_{\mathbf{k_2}}a^{*}_{\mathbf{k_3}}.
\end{equation}
Now we can extract from the sum the diagonal part $\mathbf{k_1}= \mathbf{k_3}$ (and $\mathbf{k_2}= \mathbf{k_3}$ since the system is symmetric changing the two incident gravitons), getting
\begin{equation}
    i \dot{a}_{\mathbf{k}} = \tilde\omega_{\mathbf{k}}a_{\mathbf{k}} + \sum_{\mathbf{k_1},\mathbf{k_2} \neq\mathbf{k_3}}W^{\mathbf{k_1},\mathbf{k_2}}_{\mathbf{k_3},\mathbf{k}}\delta^{\mathbf{k_1},\mathbf{k_2}}_{\mathbf{k_3},\mathbf{k}} a_{\mathbf{k_1}}a_{\mathbf{k_2}}a^{*}_{\mathbf{k_3}},
\end{equation}
where 
\begin{equation}
    \tilde\omega_{\mathbf{k}} = \omega_{\mathbf{k}} + \omega_{\rm{S},\,\mathbf{k}}\,,
\end{equation}
with 
\begin{equation}
\phantomsection
\label{omegas}
    \omega_{\rm{S},\, \mathbf{k}} = 2 \sum_{\mathbf{k_1}} W^{\mathbf{k_1},\mathbf{k}}_{\mathbf{k_1},\mathbf{k}} |a_{\mathbf{k_1}}|^2,
\end{equation}
represents a shift over the linear frequency of the system.
To eliminate the linear oscillatory term in the equation of motion, a suitable substitution may be introduced
\begin{equation}
    b_{\mathbf{k}} = \frac{a_{\mathbf{k}}e^{i\omega_{\mathbf{k}}t+ i \int\omega_{\rm{S},\mathbf{k}}dt}}{\epsilon},
\end{equation}
where $\epsilon$ is the perturbative parameter of the theory.
A simple notation in which $\mathbf{1}$, $\mathbf{2}$ and $\mathbf{3}$ coincide with $\mathbf{k_1}$, $\mathbf{k_2}$ and $\mathbf{k_3}$, respectively, has been used.\\
We obtain 
\begin{equation}
    i \dot{b}_{\mathbf{k}} = \sum_{\mathbf{1},\mathbf{2}\neq \mathbf{3}}W^{\mathbf{1},\mathbf{2}}_{\mathbf{3},\mathbf{k}}\delta^{\mathbf{1},\mathbf{2}}_{\mathbf{3},\mathbf{k}} b_{\mathbf{1}}b_{\mathbf{2}} b^{*}_{\mathbf{3}}\;e^{i\int\omega^{\mathbf{3},\mathbf{k}}_{\mathbf{1},\mathbf{2}} dt},
\end{equation}
where
\begin{equation}
\phantomsection
\label{omega123}
    \omega^{\mathbf{3} ,\mathbf{k}}_{\mathbf{1}, \mathbf{2}} = \omega_{\mathbf{3}}+\omega_{\rm{S}, \mathbf{3}} +\omega_{\mathbf{k}}+\omega_{\rm S, \mathbf{k}}-(\omega_{\mathbf{1}}+\omega_{\rm S,\mathbf{1}}+\omega_{\mathbf{2}}+\omega_{\rm S, \mathbf{2}}).
\end{equation}
Now $b_{\mathbf{k}}$ can be expanded in spherical harmonics
\begin{eqnarray}
    b_{\mathbf{k}}&=& \sum_{\ell m}b_{k,\ell m }Y_{\ell m}(\Omega_k), \\
    b_{k,\ell m}&=& \int d\Omega_k b_{\mathbf{k}}Y^*_{\ell m }(\Omega_k).
\end{eqnarray}
From this, it follows that
\begin{equation}
    i \sum_{\ell m}\dot{b}_{k,\ell m}Y_{\ell m}(\Omega_k) = \epsilon^2\sum_{\substack{\ell_i,m_i\\i=1,2,3}}\sum_{\mathbf{1},\mathbf{2}\neq\mathbf{3}}W^{\mathbf{1},\mathbf{2}}_{\mathbf{3},\mathbf{k}}\delta^{\mathbf{1},\mathbf{2}}_{\mathbf{3},\mathbf{k}} b_{1,\ell_1 m_1}b_{2,\ell_2 m_2}b^{*}_{3,\ell_3 m_3}Y_{\ell_1 m_1}(\Omega_1)Y_{\ell_2 m_2}(\Omega_2)Y^*_{\ell_3 m_3}(\Omega_3)\;e^{i\int\omega^{\mathbf{3},\mathbf{k}}_{\mathbf{1},\mathbf{2}} dt}.
\end{equation}
By projecting onto the multipoles ($\ell\, m $), the following expression is obtained:
\begin{equation}
    i \dot{b}_{k,\ell m}= \epsilon^2\sum_{\mathbf{1},\mathbf{2}\neq \mathbf{3}}\int d\Omega_k\sum_{\substack{\ell_i,m_i\\i=1,2,3}}W^{\mathbf{1},\mathbf{2}}_{\mathbf{3},\mathbf{k}}\delta^{\mathbf{1},\mathbf{2}}_{\mathbf{3},\mathbf{k}} b_{1,\ell_1 m_1}b_{2,\ell_2 m_2} b^{*}_{3,\ell_3 m_3}Y_{\ell_1 m_1}(\Omega_1)Y_{\ell_2 m_2}(\Omega_2)Y^*_{\ell_3 m_3}(\Omega_3)Y^*_{\ell m}(\Omega_k)\;e^{i\int\omega^{\mathbf{3},\mathbf{k}}_{\mathbf{1},\mathbf{2}} dt},
\end{equation}
where the orthogonality property has been used
\begin{eqnarray}
    \int d\Omega\,Y_{\ell m}(\Omega)Y^*_{\ell' m'}(\Omega)=\delta_{\ell\ell'}\delta_{mm'}.
\end{eqnarray}
The discrete sum is replaced by integrals
\begin{equation}
    \sum_{\mathbf{1},\mathbf{2} \neq \mathbf{3}} \to \prod_{i=1}^3\int d\Omega_i\int dk_i\; k_i^2  f(k_1,k_2,k_3),
\end{equation}
where 
\begin{eqnarray}\label{fdiff}
f(k_1,k_2,k_3) = 
\begin{cases}
0\;, \quad k_1,k_2 = k_3,  \\
1\;, \quad  k_1,k_2 \neq k_3.   \\
\end{cases}
\end{eqnarray}
Finally we get 
\begin{eqnarray}
    i \dot{b}_{k,\ell m}&=& \epsilon^2\sum_{\substack{\ell_i,m_i\\i=1,2,3}}\int d\Omega_k \prod_{i=1}^3\int d\Omega_i\int dk_i\; k_i^2 \; W^{\mathbf{1},\mathbf{2}}_{\mathbf{3},\mathbf{k}}\delta^{\mathbf{1},\mathbf{2}}_{\mathbf{3},\mathbf{k}} b_{1,\ell_1 m_1}b_{2,\ell_2 m_2} b^{*}_{3,\ell_3 m_3}\nonumber\\
    &\cdot& Y_{\ell_1 m_1}(\Omega_1)Y_{\ell_2 m_2}(\Omega_2)Y^*_{\ell_3 m_3}(\Omega_3)Y^*_{\ell m}(\Omega_k)\;e^{i\int\omega^{\mathbf{3},\mathbf{k}}_{\mathbf{1},\mathbf{2}} dt}f(k_1,k_2,k_3).
    \label{eom}
\end{eqnarray}
It is possible now to expand the Dirac delta and the scattering amplitude in the same harmonic basis.\\
By definition
\begin{equation}
 \delta_{\substack{k_1,\ell_1,m_1\\k_2,\ell_2,m_2}}^{\substack{k,\ell,m\\k_3,\ell_3,m_3}}=
 \dijab{1}{2}{k}{3}= \int d\Omega_k \int d\Omega_1\int d\Omega_2\int d\Omega_3\; \delta(\mathbf{k_3}+\mathbf{k}-\mathbf{k_1}-\mathbf{k_2})Y^*_{\ell m}(\Omega_k)Y^*_{\ell_3 m_3}(\Omega_3)Y_{\ell_1 m_1}(\Omega_1)Y_{\ell_2 m_2}(\Omega_2).
\end{equation}
Using the integral Dirac delta representation we can write
\begin{equation}
\dijab{1}{2}{k}{3}=
\int d\Omega_k \int d\Omega_1\int d\Omega_2\;\int d\Omega_3\; \int d^3\mathbf{x}\;  e^{i(\mathbf{k_3}+\mathbf{k}-\mathbf{k_1}-\mathbf{k_2})\mathbf{x}}Y^*_{\ell m}(\Omega_k)Y^*_{\ell_3 m_3}(\Omega_3) Y_{\ell_1 m_1}(\Omega_1)Y_{\ell_2 m_2}(\Omega_2).
\end{equation}
Recalling that
\begin{equation}
    e^{i\mathbf{k}\mathbf{x}}=4\pi\sum_{\ell m}i^{\ell}j_{\ell}(kx)Y^*_{\ell m}(\Omega_{x})Y_{\ell m}(\Omega_{k}),
\end{equation}
we obtain
\begin{eqnarray}
    \dijab{1}{2}{k}{3}& =& \sum_{\substack{\ell',m', \ell_i',m_i' \\ i=1,2,3}}\int d\Omega_k \int d\Omega_1\int d\Omega_2\int d\Omega_3 \int d\Omega_x\int dx \;x^2\; Y^*_{\ell_3 m_3}(\Omega_3) Y^*_{\ell m}(\Omega_k)Y_{\ell_1 m_1}(\Omega_1)Y_{\ell_2 m_2}(\Omega_2)\nonumber \\
    &\cdot&j_{\ell'}(k x)Y^*_{\ell' m'}(\Omega_{x})Y_{\ell' m'}(\Omega_k)j_{\ell_3'}(k_3 x)Y^*_{\ell_3' m_3'}(\Omega_{x})Y_{\ell_3' m_3'}(\Omega_3)j_{\ell_1'}(k_1 x)Y_{\ell_1' m_1'}(\Omega_{x})Y^*_{\ell_1' m_1'}(\Omega_1) \nonumber \\ &\cdot&j_{\ell_2'}(k_2 x)Y_{\ell_2' m_2'}(\Omega_{x})Y^*_{\ell_2' m_2'}(\Omega_2).
\end{eqnarray}
Consequently, the Clebsch--Gordan coefficient is defined as
\begin{equation}
    \int d\Omega_x Y_{\ell m}(\Omega_{x})Y_{\ell_1 m_1}(\Omega_{x})Y_{\ell_2 m_2}(\Omega_{x}) = \mathcal{C}_{\ell,\ell_1,\ell_2,m,m_1,m_2}\,,
\end{equation}
which gives the selection rule $|\ell_1-\ell_2|\le\ell\le \ell_1+\ell_2$.
At the end the following expression is obtained
\begin{eqnarray}
    \dijab{1}{2}{}{3}
    & =&  \int dx \;x^2 j_{\ell}(kx)j_{\ell_1}(k_1x)j_{\ell_2}(k_2x)j_{\ell_3}(k_3x)\int d\Omega_xY^*_{\ell m}(\Omega_{x})Y^*_{\ell_3 m_3}(\Omega_{x}) Y_{\ell_1 m_1}(\Omega_{x})Y_{\ell_2 m_2}(\Omega_{x}).
\end{eqnarray}
Now we can expand each couple of harmonics in the spherical harmonic basis. For example
\begin{equation}
    Y_{\ell_1 m_1}Y_{\ell_2 m_2} = \sum_{L_{12},M_{12}} A_{L_{12}M_{12}}^{\substack{\ell_1 m_1 \\ \ell_2 m_2}} Y_{L_{12}M_{12}},
\end{equation}
where 
\begin{equation}
    A_{L_{12}M_{12}}^{\substack{\ell_1 m_1 \\ \ell_2 m_2}} = \int d\Omega_x Y_{\ell_1 m_1}Y_{\ell_2 m_2}Y^*_{L_{12}M_{12}} =\mathcal{C}_{L_{12},\ell_1, \ell_2, M_{12},m_1, m_2},
\end{equation}
implying that $L_{12}= \ell_1+\ell_2$ in the eikonal limit ($\ell\gg1$).
Using the orthogonality of the spherical harmonics, we easily get 
\begin{equation}    
\dijab{1}{2}{}{3}= \sum_{L_{12},M_{12}}\int d\Omega_x \;\mathcal{C}_{L_{12},\ell_1,\ell_2,M_{12},m_1,m_2} Y_{\ell m }^*(\Omega_{x})Y^*_{\ell_3 m_3 }(\Omega_{x})Y_{L_{12} M_{12}}(\Omega_{x})\mathcal{I}(\ell,\ell_1,\ell_2,\ell_3,k,k_1,k_2,k_3),
\end{equation}
where 
\begin{equation}
    \mathcal{I}(\ell,\ell_1,\ell_2,\ell_3,k,k_1,k_2,k_3) = \int dx\; x^2 j_{\ell}(kx)j_{\ell_1}(k_1x)j_{\ell_2}(k_2x)j_{\ell_3}(k_3x),
\end{equation}
ensure momentum conservation \cite{RMehrem_1991}. We can finally write 
\begin{equation}
    \dijab{1}{2}{}{3}= \sum_{L_{12},M_{12}} \;\mathcal{C}_{L_{12},\ell_1,\ell_2,M_{12},m_1,m_2} \mathcal{C}_{L_{12},\ell,\ell_3,M_{12},m,m_3}\mathcal{I}(\ell,\ell_1,\ell_2,\ell_3,k,k_1,k_2,k_3) ,
\end{equation}
that in the eikonal limit ensures $\ell_1+\ell_2 = \ell_3 +\ell$.
Therefore we can expand the Dirac delta in the following way
\begin{equation}
    \delta(\mathbf{k_3}+\mathbf{k}-\mathbf{k_1}-\mathbf{k_2}) = \sum_{\substack{\ell,m,\ell_i,m_i \\ i=1,2,3}}  
    \dijab{1}{2}{k}{3}Y_{\ell m}(\Omega_k)Y_{\ell_3 m_3}(\Omega_3)Y^*_{\ell_1 m_1}(\Omega_1)Y^*_{\ell_2 m_2}(\Omega_2).
\end{equation}
We can expand the scattering amplitude as well,
\begin{eqnarray}
 \Wijab{1}{2}{}{3}&=&\int d\Omega_k\int d\Omega_{1}\int d\Omega_{2}\int d\Omega_{3}\,W^{\mathbf{1},\mathbf{2}}_{\mathbf{3},\mathbf{k}}\,Y^*_{\ell m}(\Omega_k)Y^*_{\ell_3 m_3}(\Omega_3)Y_{\ell_1 m_1}(\Omega_1)Y_{\ell_2 m_2}(\Omega_2).
    \end{eqnarray}
Consequently the amplitude can be written as
\begin{equation}
    W^{\mathbf{1},\mathbf{2}}_{\mathbf{3},\mathbf{k}} = \sum_{\substack{\ell,m,\ell_i,m_i \\ i=1,2,3}}  
    \Wijab{1}{2}{}{3}Y_{\ell m}(\Omega_k)Y_{\ell_3 m_3}(\Omega_3)Y^*_{\ell_1 m_1}(\Omega_1)Y^*_{\ell_2 m_2}(\Omega_2).
\end{equation}
Replacing this two expansions in the equation of motion Eq. (\ref{eom}), we calculate 
\begin{eqnarray}
     i \dot{b}_{k,\ell m} &=& \epsilon^2\sum_{\substack{\ell_i,m_i \\ i=1,2,3}} \sum_{\substack{\tilde{\ell} ,\tilde{m},\tilde{\ell}_i,\tilde{m}_i\\ i=1,2,3}}\sum_{\substack{\ell', m', \ell'_i,m'_i\\ i=1,2,3}}\int d\Omega_k \prod_{i=1}^3\int dk_i k_i^2 \int d\Omega_i\ Y^*_{\ell m}(\Omega_k) Y^*_{\ell_3 m_3}(\Omega_3)Y_{\ell_1 m_1}(\Omega_1)Y_{\ell_2 m_2}(\Omega_2)
     \nonumber\\
     &\cdot& Y_{\ell' m'}(\Omega_k) Y_{\ell_3' m_3'}(\Omega_3)Y^*_{\ell'_1 m'_1}(\Omega_1)Y^*_{\ell'_2 m'_2}(\Omega_2) Y_{\tilde\ell \tilde m}(\Omega_k) Y_{\tilde{\ell}_3 \tilde {m}_3}(\Omega_3)Y^*_{\tilde{\ell}_1 \tilde {m}_1}(\Omega_1)Y^*_{\tilde{\ell}_2 \tilde {m}_2}(\Omega_2)   \nonumber\\
     &\cdot& 
     \Wijab{1'}{2'}{k'}{3'}\dijab{\tilde 1}{\tilde 2}{\tilde k}{\tilde 3}b_{1,\ell_1 m_1}b_{2,\ell_2m_2}b_{3,\ell_3m_3}^{*}\;e^{i\int\omega^{3,k}_{1,2} dt}f(k_1,k_2,k_3),
\end{eqnarray}
where $[k']=(k,\ell', m')$ and $[\tilde{k}]=(k,\tilde\ell, \tilde m)$. From here 
\begin{eqnarray}
     i \dot{b}_{k,\ell m} &=& \epsilon^2\sum_{\substack{\ell_i,m_i \\ i=1,2,3}} \sum_{\substack{\tilde{\ell} ,\tilde{m},\tilde{\ell}_i,\tilde{m}_i\\ i=1,2,3}}\sum_{\substack{\ell', m', \ell'_i,m'_i\\ i=1,2,3}}\int d\Omega_k \prod_{i=1}^3\int dk_i k_i^2 \mathcal{C}_{\ell,\ell',\tilde\ell,m,m',\tilde m} \mathcal{C}_{\ell_i,\ell'_i,\tilde{\ell}_i,m_i,m'_i,\tilde {m}_i} \nonumber\\
     &\cdot& 
      \Wijab{1'}{2'}{k'}{3'}\dijab{\tilde 1}{\tilde 2}{\tilde k}{\tilde 3}b_{1,\ell_1 m_1}b_{2,\ell_2 m_2}b_{3,\ell_3 m_3}^{*}\;e^{i\int\omega^{3,k}_{1,2} dt}f(k_1,k_2,k_3).
\end{eqnarray}
Using the properties of the Clebsch-Gordan coefficients, we write
\begin{equation}
\begin{cases}
\tilde\ell= \ell-\ell' &  \\
\tilde\ell_i= \ell_i-\ell_i'. 
\end{cases}
\end{equation}
It is trivial to see
\begin{eqnarray}
     i \dot{b}_{k,\ell m} &=& \epsilon^2\sum_{\substack{\ell_i,m_i \\ i=1,2,3}} \sum_{\substack{\ell', m', \ell'_i,m'_i\\ i=1,2,3}}\prod_{i=1}^3\int dk_i k_i^2 \mathcal{C}_{\ell,\ell',\ell-\ell',m,m',m-m'} \mathcal{C}_{\ell_i,\ell'_i,\ell_i-\ell'_i,m_i,m'_i,m_i-m'_i} \nonumber\\
     &\cdot& 
     \Wijab{1'}{2'}{k'}{3'}\dijab{1-1'}{2-2'}{k-k'}{3-3'}b_{1,\ell_1 m_1}b_{2,\ell_2 m_2}b_{3,\ell_3 m_3}^{*}\;e^{i\int\omega^{3,k}_{1,2} dt}f(k_1,k_2,k_3),
\end{eqnarray}
where $[k-k']=(k,\ell-\ell', m-m')$ .We can generalize it in the following way 
\begin{eqnarray}
     i \dot{b}_{k,\ell m} &=& \epsilon^2\sum_{\substack{\ell_i,m_i \\ i=1,2,3}} \prod_{i=1}^3\int dk_i k_i^2 
     \LA b_{1,\ell_1 m_1}b_{2,\ell_2 m_2}b_{3,\ell_3 m_3}^{*}\;e^{i\int\omega^{3,k}_{1,2} dt} f (k_1,k_2,k_3),
\end{eqnarray}
where 
\begin{eqnarray}\label{Aterm}
    \LA&=&\mathcal{A}_{\substack{k_1,\ell_1,m_1\\k_2,\ell_2,m_2}}^{\substack{k,\ell,m\\k_3,\ell_3,m_3}} = \sum_{\substack{\ell', m' \ell'_i,m'_i\\ i=1,2,3}} 
     \Wijab{1'}{2'}{k'}{3'}\dijab{1-1'}{2-2'}{k-k'}{3-3'}
    \mathcal{C}_{\ell,\ell',\ell-\ell',m,m',m-m'} \nonumber\\
     &\cdot& \mathcal{C}_{\ell_1,\ell'_1,\ell_1-\ell'_1,m_1,m'_1,m_1-m'_1}\mathcal{C}_{\ell_2,\ell'_2,\ell_2-\ell'_2,m_2,m'_2,m_2-m'_2}\mathcal{C}_{\ell_3,\ell'_3,\ell_3-\ell'_3,m_3,m'_3,m_3-m'_3},
\end{eqnarray}
is the convolution in the harmonic space of the scattering amplitude with the delta function and  the Clebsch-Gordan coefficients. It is now necessary to expand \(b_{k,\ell m}\) around a time \(T\), lying between the characteristic linear timescale \(\tau_{\rm{L}} = 2\pi / \omega_{k}\) and the timescale at which nonlinear effects become significant, \(\tau_{\rm{NL}} = 2\pi / (\epsilon^{4}\omega_{k})\).
Considering $\tau_{\rm{L}}\ll T\ll \tau_{\rm{NL}}$
\begin{equation}\label{expT}
    b_{k,\ell m}(T)= b_{k,\ell m }^{(0)} + \epsilon^2 b_{k,\ell m }^{(1)}+\epsilon^4 b_{k,\ell m }^{(2)}+ O(\epsilon^6),
\end{equation}
where $ b_{k,\ell m }^{(0)}$ is a constant background for which $ b_{k,\ell m }^{(0)}=  b_{k,\ell m }^{(0)}(t=0)$. At first order in $\epsilon$, 
\begin{eqnarray}
    i \epsilon^2 \dot{b}_{k,\ell m}^{(1)} = \epsilon^2 \sum_{\substack{\ell_i,m_i \\ i=1,2,3}}\prod_{i=1}^3\int dk_i k_i^2 
    \LA b_{1,\ell_1 m_1}^{(0)}b_{2,\ell_2,m_2}^{(0)}b_{3,\ell_3,m_3}^{*(0)}\;e^{i\int\omega^{3,k}_{1,2} dt} f (k_1,k_2,k_3).
\end{eqnarray}
Since \(\omega^{3,k}_{1,2}\) acquires its time dependence solely from the 
presence of the off-diagonal shift \(\omega_{\mathrm{S}}(t)\), 
of order \(O(\epsilon^{2})\), while  the linear contribution 
\(\omega_{k}\) scales as \(O(\epsilon^{0})\), we can write

\begin{eqnarray}
     b_{k,\ell m}^{(1)} =-i  \sum_{\substack{\ell_i,m_i \\ i=1,2,3}}\prod_{i=1}^3\int dk_i k_i^2 
     \LA b_{1,\ell_1 m_1}^{(0)}b_{2,\ell_2m_2}^{(0)}b_{3,\ell_3m_3}^{*(0)} f (k_1,k_2,k_3)\int_0^T dt \;e^{i\omega^{3,k}_{1,2}t },
\end{eqnarray}
where we call
\begin{eqnarray}
    \Delta(\omega^{3k}_{12}) = \int_0^T dt \;e^{i\omega^{3,k}_{1,2} t} = \frac{e^{i\omega^{3,k}_{1,2}T}-1}{i \omega^{3,k}_{1,2}}.
\end{eqnarray}
The first order will be
\begin{eqnarray}
     b_{k,\ell m}^{(1)} =-i  \sum_{\substack{\ell_i,m_i \\ i=1,2,3}}\prod_{i=1}^3\int dk_i k_i^2 
     \LA b_{1,\ell_1 m_1}^{(0)}b_{2,\ell_2m_2}^{(0)}b_{3,\ell_3m_3}^{*(0)} f (k_1,k_2,k_3)\Delta(\omega^{3k}_{12}).
\end{eqnarray}
Performing all the calculations for the second order, we obtain
\begin{eqnarray}\label{2expb}
    b_{k,\ell m}^{(2)} &= &-\sum_{\substack{\ell_i,m_i \\ i=1,2,3}}\sum_{\substack{\ell_n,m_n \\ n=4,5,6}}\prod_{i=1}^3\prod_{j=4}^6\int dk_i k_i^2\int dk_j k_j^2 
\LA \Aijab{4}{3}{1}{6}
b_{2,\ell_2m_2}^{(0)}b_{3,\ell_3m_3}^{*(0)} \nonumber\\
&\cdot&b_{4,\ell_4m_4}^{(0)}b_{5,\ell_5m_5}^{(0)}b_{6,\ell_6m_6}^{*(0)}\Delta(\omega^{61}_{45})\Delta(\omega^{3k}_{12})f(k_1,k_2,k_3)f(k_4,k_5,k_6)  \nonumber\\
     &-&
     \sum_{\substack{\ell_i,m_i \\ i=1,2,3}}\sum_{\substack{\ell_n,m_n \\ n=4,5,6}}\prod_{i=1}^3\prod_{j=4}^6\int dk_i k_i^2\int dk_j k_j^2 
 \LA \Aijab{4}{5}{2}{6} b_{1,\ell_1m_1}^{(0)}b_{3,\ell_3m_3}^{*(0)} \nonumber\\
&\cdot&b_{4,\ell_4m_4}^{(0)}b_{5,\ell_5m_5}^{(0)}b_{6,\ell_6m_6}^{*(0)}\Delta(\omega^{62}_{45})\Delta(\omega^{3k}_{12})f(k_1,k_2,k_3)f(k_4,k_5,k_6)  \nonumber\\
     &+&\frac{1}{2}
     \sum_{\substack{\ell_i,m_i \\ i=1,2,3}}\sum_{\substack{\ell_n,m_n \\ n=4,5,6}}\prod_{i=1}^3\prod_{j=4}^6\int dk_i k_i^2\int dk_j k_j^2 
    \LA \left(\Aijab{4}{5}{3}{6}\right)^* b_{1,\ell_1m_1}^{(0)}b_{2,\ell_2m_2}^{(0)} \nonumber\\
     &\cdot&b_{4,\ell_4m_4}^{*(0)}b_{5,\ell_5m_5}^{*(0)}b_{6,\ell_6m_6}^{(0)}\Delta^*(\omega^{63}_{45})\Delta(\omega^{3k}_{12})f(k_1,k_2,k_3)f(k_4,k_5,k_6).
\end{eqnarray}
As mentioned in the text, following  Ref. \cite{t12}, we treat \( b_{k,\ell m} \) as a stochastic variable, writing it as \( b_{k_i,\ell_i m_i} = \sqrt{J_{k_i,\ell_i m_i}} \, e^{i\phi_i} \), where \( J_{k_i,\ell_im_i} \in \mathbb{R}^+ \) is the intensity and \( \phi_i \) is the phase. We assume uncorrelated distributions for the intensity and phase of each mode, and that phases of different modes are also uncorrelated. It is necessary to  define the one-mode generative function calculated at the intermediate time $T$
\begin{eqnarray}
    \mathcal{L}_{k,\ell m }(\lambda_{k,\ell m },T) = \langle e^{\lambda_{k,\ell m }|b_{k,\ell m}(T)|^2}\rangle,
\end{eqnarray}
where $|b_{k,\ell m}(T)|^2 = J_{k,\ell m}$. Substituting the expansion (\ref{expT}) and, after some algebra, we obtain 
\begin{eqnarray}
\mathcal{L}_{k,\ell m }(\lambda_{k,\ell m },T) 
&=&\Bigg\langle 
e^{\lambda_{k,\ell m }|b_{k,\ell,m}^{(0)}|^2}
\Bigg[
1 + \lambda_{k,\ell m } \epsilon^2 
\left(b_{k,\ell m}^{* (0)}b_{k,\ell m}^{(1)} + \mathrm{c.c.}\right) 
+ \lambda_{k,\ell m } \epsilon^4
\Bigg(
|b_{k,\ell m}^{(1)}|^2
+ \left(b_{k,\ell m}^{*(0)}b_{k,\ell m}^{(2)} + \mathrm{c.c.}\right)\nonumber\\
&+& \frac{\lambda_{k,\ell m }}{2}
\bigg[\left(\left(b_{k,\ell m}^{* (0)}b_{k,\ell m}^{(1)}\right)^2 + \mathrm{c.c.}\right)
+2 |b_{k,\ell m}^{* (0)}|^2|b_{k,\ell m}^{ (1)}|^2\bigg]
\Bigg)
\Bigg]
\Bigg\rangle.
\end{eqnarray}
We can notice that ${\cal O}(\epsilon^0)$ just gives $\mathcal{L}_{k,\ell m }(\lambda_{k,\ell m },0)$. Since the modes 
\(b_{k,\ell m}\) consists of an intensity component and a phase factor, an average must be taken over the respective distributions.\\
At order $\epsilon^4$ follows
\begin{eqnarray}\label{generativeL}
\mathcal{L}_{k,\ell m }(\lambda_{k,\ell m },T)
- \mathcal{L}_{k,\ell m }(\lambda_{k,\ell m },0)
&\simeq&
\epsilon^4 \bigg\langle
e^{\lambda_{k,\ell m }|b_{k,\ell m}^{(0)}|^2}
\lambda_{k,\ell m }
\Big(
\langle |b_{k,\ell m}^{(1)}|^2 \rangle_{\phi}
+ \lambda_{k,\ell m } J_{k,\ell m}
\langle |b_{k,\ell m}^{(1)}|^2 \rangle_{\phi}
\Big)
\nonumber\\
&+& \lambda_{k,\ell m }
\big\langle
b_{k,\ell m}^{*(0)} b_{k,\ell m}^{(2)} + \mathrm{c.c.}
\big\rangle_{\phi}
\bigg\rangle_{J_{k,\ell m}}.
\end{eqnarray}
Here we have already assumed that $\langle b_{k,\ell m}^{* (0)}b_{k,\ell m}^{(1)}\rangle_{\phi} = 0$ and $\Big\langle \left(b_{k,\ell m}^{* (0)}b_{k,\ell m}^{(1)}\right)^2\Big\rangle_{\phi} = 0$ as we will demonstrate in the following lines.
The next step is to compute the average over the random phases, which are assumed to be uniformly distributed. For the first correlator, 
\begin{eqnarray}
    \langle b_{k,\ell m}^{* (0)}b_{k,\ell m}^{(1)}\rangle_{\phi} = -i  \sum_{\substack{\ell_i,m_i \\ i=1,2,3}}\prod_{i=1}^3\int dk_i k_i^2 
    \LA \langle b_{k,\ell m}^{*(0)} b_{1,\ell_1 m_1}^{(0)}b_{2,\ell_2m_2}^{(0)}b_{3,\ell_3m_3}^{*(0)}\rangle_{\phi} f (k_1,k_2,k_3)\Delta(\omega^{3k}_{12}),
\end{eqnarray}
using the Wick's rule contraction, it is possible to  contract just $b^{*}_{k,\ell m}$ with $b_{1,\ell_1 m_1}$ and $b_{2,\ell_2 m_2}$ with $b^{*}_{3,\ell_3 m_3}$ (i.e. $k1,23$)  or $b^{*}_{k,\ell m}$ with $b_{2,\ell_2 m_2}$, and $b_{1,\ell_1 m_1}$ with $b^{*}_{3,\ell_3 m_3}$ (i.e. $k2,13$). Nevertheless, since $k_1$ and $k_2$ should be different from $k_3$, the two contributions will be zero. A similar reasoning could be applied to show that $\Big\langle \left(b_{k,\ell m}^{* (0)}b_{k,\ell m}^{(1)}\right)^2\Big\rangle_{\phi} = 0$. Now we have to calculate
\begin{eqnarray}
     \langle b_{k,\ell m}^{* (1)}b_{k,\ell m}^{(1)}\rangle_{\phi} &=&\sum_{\substack{\ell_i,m_i \\ i=1,2,3}}\sum_{\substack{\ell_n,m_n \\ n=4,5,6}}\prod_{i=1}^3\prod_{j=4}^6\int dk_i k_i^2\int dk_j k_j^2 
     \LA \left(\Aijab{4}{5}{k}{6}\right)^*
     f(k_1,k_2,k_3)f(k_4,k_5,k_6) \nonumber \\ &\cdot&\langle b_{1,\ell_1 m_1}^{(0)}b_{2,\ell_2m_2}^{(0)}b_{3,\ell_3m_3}^{*(0)}b_{4,\ell_4 m_4}^{*(0)}b_{5,\ell_5m_5}^{*(0)}b_{6,\ell_6m_6}^{(0)}\rangle_{\phi}\Delta(\omega^{3k}_{12})\Delta^*(\omega^{6k}_{45}).
\end{eqnarray}
It is easy to notice that the only non-zero contributions are $(41, 52, 63)$ or $(51,42,63)$ (which is symmetric switching the two incoming gravitons $4\leftrightarrow 5$). By definition,
\begin{eqnarray}
    \langle b_{1,\ell_1 m_1}^{(0)}b_{2,\ell_2m_2}^{(0)}b_{3,\ell_3m_3}^{*(0)}b_{4,\ell_4 m_4}^{*(0)}b_{5,\ell_5m_5}^{*(0)}b_{6,\ell_6m_6}^{(0)}\rangle_{\phi} &=& 2 \frac{\delta_{41}}{k_1^2}\delta_{\ell_4\ell_1}\delta_{m_4m_1}\frac{\delta_{52}}{k_2^2}\delta_{\ell_5\ell_2}\delta_{m_5m_2}\frac{\delta_{63}}{k_3^2}\delta_{\ell_6\ell_3}\delta_{m_6m_3}\nonumber \\ &\cdot&J_{1,\ell_1 m_1}J_{2,\ell_2 m_2}J_{3,\ell_3 m_3},
\end{eqnarray}
it follows
\begin{eqnarray}
    \langle b_{k,\ell m}^{* (1)}b_{k,\ell m}^{(1)}\rangle_{\phi} &=& 2 \sum_{\substack{\ell_i,m_i \\ i=1,2,3}}\prod_{i=1}^3\int dk_i k_i^2\left|
    \LA 
    \right|^2\left|\Delta(\omega^{3k}_{12}) \right|^2 f(k_1,k_2,k_3) J_{1,\ell_1 m_1}^{(0)}J_{2,\ell_2m_2}^{(0)}J_{3,\ell_3m_3}^{(0)}.
\end{eqnarray}
Now we should calculate the last angular correlator $\langle  b_{k,\ell m}^{*(0)} b_{k,\ell m}^{(2)}\rangle_{\phi}$.
Performing all the calculations we obtain
\begin{eqnarray}
     \langle  b_{k,\ell m}^{*(0)} b_{k,\ell m}^{(2)}+\rm{c.c} \rangle_{\phi} &=& -4 \sum_{\substack{\ell_i,m_i \\ i=1,2,3}}\prod_{i=1}^3\int dk_i k_i^2 \left|
     \LA 
\right|^2\mathcal{R}\left[E(\omega^{3k}_{12},\omega^{12}_{3k})\right]f(k_1,k_2,k_3)\nonumber \\ &\cdot& J_{k,\ell m}^{(0)}J_{2,\ell_2 m_2}^{(0)}J_{3,\ell_3 m_3}^{(0)} \nonumber\\ &-4&
     \sum_{\substack{\ell_i,m_i \\ i=1,2,3}}\prod_{i=1}^3\int dk_i k_i^2 \left|
     \LA 
\right|^2\mathcal{R}\left[E(\omega^{3k}_{12},\omega^{12}_{3k})\right]f(k_1,k_2,k_3)\nonumber \\ &\cdot& J_{k,\ell m}^{(0)}J_{1,\ell_1 m_1}^{(0)}J_{3,\ell_3 m_3}^{(0)} \nonumber \\ &+2& \sum_{\substack{\ell_i,m_i \\ i=1,2,3}}\prod_{i=1}^3\int dk_i k_i^2\left|
\LA \right|^2 \left|\Delta(\omega^{3k}_{12})\right|^2 f(k_1,k_2,k_3)\nonumber\\ &\cdot& J_{k,\ell  m}^{(0)}J_{1,\ell_1m_1}^{(0)}J_{2,\ell_2m_2}^{(0)},
\end{eqnarray}
where we have defined
\begin{equation}
    E(x,-x) = \Delta(x)\Delta(-x)
\end{equation}
and we have used the symmetric properties of the scattering amplitude $W ^{12}_{3k} = (W^{3k}_{12})^*$ and $W ^{12}_{3k} = W^{21}_{k3}$.\\
We can substitute all this contraction in Eq. (\ref{generativeL}). The whole equation reads
\begin{eqnarray}
    &&\mathcal{L}_{k,\ell m }(\lambda_{k,\ell m },T)
- \mathcal{L}_{k,\ell m }(\lambda_{k,\ell m },0)= 2\epsilon^4 \left(\mathcal{L}_{k,\ell m}\lambda_{k,\ell m }+\frac{\partial \mathcal{L}_{k,\ell m}}{\partial\lambda_{k,\ell m}}\lambda^2_{k,\ell m }\right)\nonumber \\ &&\cdot
\sum_{\substack{\ell_i,m_i \\ i=1,2,3}}\prod_{i=1}^3\int dk_i k_i^2\left|
\LA \right|^2\left|\Delta(\omega^{3k}_{12}) \right|^2 f(k_1,k_2,k_3) n_{1,\ell_1 m_1}n_{2,\ell_2m_2}n_{3,\ell_3m_3}\nonumber\\ &&-4\epsilon^4 \lambda_{k,\ell m }\frac{\partial \mathcal{L}_{k,\ell m}}{\partial\lambda_{k,\ell m}}\sum_{\substack{\ell_i,m_i \\ i=1,2,3}}\prod_{i=1}^3\int dk_i k_i^2 \left|
\LA \right|^2\mathcal{R}\left[E(\omega^{3k}_{12},\omega^{12}_{3k})\right]f(k_1,k_2,k_3) n_{2,\ell_2 m_2}n_{3,\ell_3 m_3} \nonumber\\ &&-4\epsilon^4\lambda_{k,\ell m }\frac{\partial \mathcal{L}_{k,\ell m}}{\partial\lambda_{k,\ell m}}
     \sum_{\substack{\ell_i,m_i \\ i=1,2,3}}\prod_{i=1}^3\int dk_i k_i^2 \left|
\LA \right|^2\mathcal{R}\left[E(\omega^{3k}_{12},\omega^{12}_{3k})\right]f(k_1,k_2,k_3) n_{1,\ell_1 m_1}n_{3,\ell_3 m_3} \nonumber \\ &&+2\epsilon^4\lambda_{k,\ell m }\frac{\partial \mathcal{L}_{k,\ell m}}{\partial\lambda_{k,\ell m}} \sum_{\substack{\ell_i,m_i \\ i=1,2,3}}\prod_{i=1}^3\int dk_i k_i^2\left|
\LA \right|^2 \left|\Delta(\omega^{3k}_{12})\right|^2 f(k_1,k_2,k_3) n_{1,\ell_1m_1}n_{2,\ell_2m_2},
\end{eqnarray}
where we have used the following statistical properties:
\begin{eqnarray}
    \bigg\langle e^{\lambda_{k,\ell m }J^{(0)}_{k,\ell m }}J^{(0)}_{1,\ell_1 m_1 }J^{(0)}_{2,\ell_2 m_2 }J^{(0)}_{3,\ell_3 m_3 }\bigg\rangle &=& \bigg\langle J^{(0)}_{1,\ell_1 m_1 }\bigg\rangle\bigg\langle J^{(0)}_{2,\ell_2 m_2 }\bigg\rangle\bigg\langle J^{(0)}_{3,\ell_3 m_3 }\bigg\rangle\bigg\langle e^{\lambda_{k,\ell m }J^{(0)}_{k,\ell m }} \bigg\rangle=n_{1,\ell_1m_1}n_{2,\ell_2m_2}n_{3,\ell_3m_3}\mathcal{L}_{k,\ell m}\nonumber\\
    &&
\end{eqnarray}
and
\begin{eqnarray}
    \bigg\langle e^{\lambda_{k,\ell m }J^{(0)}_{k,\ell m }}J^{(0)}_{k,\ell m}J^{(0)}_{1,\ell_1 m_1 }J^{(0)}_{2,\ell_2 m_2 }J^{(0)}_{3,\ell_3 m_3 }\bigg\rangle &=& \bigg\langle J^{(0)}_{1,\ell_1 m_1 }\bigg\rangle\bigg\langle J^{(0)}_{2,\ell_2 m_2 }\bigg\rangle\bigg\langle J^{(0)}_{3,\ell_3 m_3 }\bigg\rangle\bigg\langle J^{(0)}_{k,\ell m}e^{\lambda_{k,\ell m }J^{(0)}_{k,\ell m }} \bigg\rangle\nonumber \\ &=&n_{1,\ell_1m_1}n_{2,\ell_2m_2}n_{3,\ell_3m_3}\frac{\partial \mathcal{L}_{k,\ell m}}{\partial\lambda_{k,\ell m}}.
\end{eqnarray}
The occupation number for gravitons has been defined, with a specific set of ($k,\ell,m$), as $n_{k,\ell m } = \left \langle J^{(0)}_{k,\ell m}\right \rangle$.\\
At this stage, the perturbative parameter $\epsilon$ is set to zero and, as a consequence, $T\to \infty$. It is easy to demonstrate that, in this case
\begin{eqnarray}
    &&\left|\Delta(x)\right|^2 \to 2\pi T\delta(x),  \nonumber \\ &&\mathcal{R}\left[E(x,-x)\right]\to \pi T\delta(x).
\end{eqnarray}
Therefore, the expression can be written as
\begin{eqnarray}
    &&\dot{\mathcal{L}}_{k,\ell m }(\lambda_{k,\ell m },T)
= 4\pi\epsilon^4 \left(\mathcal{L}_{k,\ell m}\lambda_{k,\ell m }+\frac{\partial \mathcal{L}_{k,\ell m}}{\partial\lambda_{k,\ell m}}\lambda^2_{k,\ell m }\right)\nonumber \\ &&\cdot
\sum_{\substack{\ell_i,m_i \\ i=1,2,3}}\prod_{i=1}^3\int dk_i k_i^2\left|
\LA \right|^2\delta(\omega^{3k}_{12})  f(k_1,k_2,k_3) n_{1,\ell_1 m_1}n_{2,\ell_2m_2}n_{3,\ell_3m_3}\nonumber\\ &&-4\pi\epsilon^4 \lambda_{k,\ell m }\frac{\partial \mathcal{L}_{k,\ell m}}{\partial\lambda_{k,\ell m}}\sum_{\substack{\ell_i,m_i \\ i=1,2,3}}\prod_{i=1}^3\int dk_i k_i^2 \left|
\LA \right|^2\delta(\omega^{3k}_{12})f(k_1,k_2,k_3) n_{2,\ell_2 m_2}n_{3,\ell_3 m_3} \nonumber\\ &&-4\pi\epsilon^4\lambda_{k,\ell m }\frac{\partial \mathcal{L}_{k,\ell m}}{\partial\lambda_{k,\ell m}}
     \sum_{\substack{\ell_i,m_i \\ i=1,2,3}}\prod_{i=1}^3\int dk_i k_i^2 \left|
\LA \right|^2\delta(\omega^{3k}_{12})f(k_1,k_2,k_3) n_{1,\ell_1 m_1}n_{3,\ell_3 m_3} \nonumber \\ &&+4\pi\epsilon^4\lambda_{k,\ell m }\frac{\partial \mathcal{L}_{k,\ell m}}{\partial\lambda_{k,\ell m}} \sum_{\substack{\ell_i,m_i \\ i=1,2,3}}\prod_{i=1}^3\int dk_i k_i^2\left|
\LA \right|^2 \delta(\omega^{3k}_{12}) f(k_1,k_2,k_3) n_{1,\ell_1m_1}n_{2,\ell_2m_2}.
\end{eqnarray}
The equation can now be generalized and expressed in a more compact form
\begin{eqnarray}
    \dot{\mathcal{L}}_{k,\ell m }(\lambda_{k,\ell m },T) = \lambda_{k,\ell m }\eta_{k,\ell m }\mathcal{L}_{k,\ell m }+\left(\lambda^2_{k,\ell m }\eta_{k,\ell m }-\lambda_{k,\ell m }\gamma_{k,\ell m }\right)\frac{\partial \mathcal{L}_{k,\ell m}}{\partial\lambda_{k,\ell m}},
\end{eqnarray}
where 
\begin{eqnarray}
    \eta_{k,\ell m} = 4\pi\epsilon^4\sum_{\substack{\ell_i,m_i \\ i=1,2,3}}\prod_{i=1}^3\int dk_i k_i^2\left|
\LA \right|^2\delta(\omega^{3k}_{12})  f(k_1,k_2,k_3) n_{1,\ell_1 m_1}n_{2,\ell_2m_2}n_{3,\ell_3m_3},
\end{eqnarray}
and
\begin{eqnarray}
    \gamma_{k,\ell m}&=&4\pi\epsilon^4\sum_{\substack{\ell_i,m_i \\ i=1,2,3}}\prod_{i=1}^3\int dk_i k_i^2\left|
\LA \right|^2\delta(\omega^{3k}_{12})  f(k_1,k_2,k_3)\nonumber\\ 
    &\cdot&\left( n_{2,\ell_2m_2}n_{3,\ell_3m_3}+n_{1,\ell_1m_1}n_{3,\ell_3m_3}-n_{1,\ell_1m_1}n_{2,\ell_2m_2}\right).
\end{eqnarray}
The general Boltzmann equation can now be written as \cite{t12}
\begin{eqnarray}
    \dot{n}_{k,\ell m}= \eta_{k,\ell m}-\gamma_{k,\ell m}n_{k,\ell m},
\end{eqnarray}
from which it follows that
\begin{eqnarray}
\label{boltzmann}
    \dot{n}_{k,\ell m} &=& 4\pi\epsilon^4\sum_{\substack{\ell_i,m_i \\ i=1,2,3}}\prod_{i=1}^3\int dk_i k_i^2\left|
\LA \right|^2\delta(\omega^{3k}_{12})  f(k_1,k_2,k_3)n_{k,\ell m}n_{1,\ell_1 m_1}n_{2,\ell_2 m_2}n_{3,\ell_3 m_3}  \nonumber\\ &\cdot&
    \left[\frac{1}{n_{k,\ell m}}+\frac{1}{n_{3,\ell_3 m_3}}-\frac{1}{n_{1,\ell_1 m_1}}-\frac{1}{n_{2,\ell_2 m_2}}\right].
\end{eqnarray}
This is the final Boltzmann equation in multipoles space. This is very similar to one one found in Ref. \cite{t12}, where it has been written only in momentum space.
\section{An index for the turbulent cascade: the eikonal limit calculation}
\noindent
It is possible to symmetrize the Boltzmann equation (\ref{boltzmann}) as follows
\begin{eqnarray}\label{boltzsymm}
\dot{n}_{k,\ell m} &=& \frac{1}{4}\sum_{\substack{\ell_i,m_i \\ i=1,2,3}}\prod_{i=1}^3\int dk_i k_i^2  n_{k,\ell m}n_{1,\ell_1 m_1}n_{2,\ell_2 m_2}n_{3,\ell_3 m_3}\left[
\Jijab{1}{2}{k}{3}+ \Jijab{1}{2}{3}{k}
-
\Jijab{k}{2}{1}{3}
-\Jijab{1}{k}{2}{3}
\right],\nonumber\\
&&
\end{eqnarray}
where
\begin{eqnarray}\label{Iterm}
\Jijab{1}{2}{k}{3}\equiv 4\pi\epsilon^4 \left|
\LA \right|^2\delta(\omega^{3k}_{12})\left[\frac{1}{n_{k,\ell m}}+\frac{1}{n_{3,\ell_3 m_3}}-\frac{1}{n_{1,\ell_1 m_1}}-\frac{1}{n_{2,\ell_2 m_2}}\right]f(k_1,k_2,k_3).
\end{eqnarray}
A change of variables is now introduced to shift each of the last three integrals into the first one. As an example, we consider the second integral term:
\begin{eqnarray}\label{secondint}
\frac{1}{4}\sum_{\substack{\ell_i,m_i \\ i=1,2,3}}\prod_{i=1}^3\int dk_i k_i^2  
\Jijab{1}{2}{3}{k}n_{k,\ell m}n_{1,\ell_1 m_1}n_{2,\ell_2 m_2}n_{3,\ell_3 m_3}. 
\end{eqnarray}
For this last term, it is possible to perform the ZT transformations \cite{Zakharov}
\begin{eqnarray}\label{coordtransf}
    &&k_1=\frac{k\tilde{k}_1}{\tilde{k}_3} \qquad k_2=\frac{k\tilde{k}_2}{\tilde{k}_3}\qquad k_3=\frac{k^2}{\tilde{k}_3},\nonumber\\
    &&\ell_1=\frac{\ell\tilde{\ell}_1}{\tilde{\ell}_3} \qquad \ell_2=\frac{\ell\tilde{\ell}_2}{\tilde{\ell}_3}\qquad \ell_3=\frac{\ell^2}{\tilde{\ell}_3},\\
    &&m_1=\frac{m\tilde{m}_1}{\tilde{m}_3} \qquad m_2=\frac{m\tilde{m}_2}{\tilde{m}_3} \qquad m_3=\frac{m^2}{\tilde{m}_3}\nonumber.
\end{eqnarray}
Defining $R_{k_3}$ and $R_{\ell_3}$ as
\begin{eqnarray}
    R_{k_3}&\equiv&\frac{k}{\tilde{k}_3}\label{tr1}\\
    R_{\ell_3}&\equiv&\frac{\ell}{\tilde{\ell}_3},\label{tr2}
\end{eqnarray}
we simply have that
\begin{eqnarray}
(k_1k_2k_3)^2dk_1dk_2dk_3\to(\tilde{k}_1\tilde{k}_2\tilde{k}_3)^2(R_{k_3})^{12}d\tilde{k}_1d\tilde{k}_2d\tilde{k}_3.
\end{eqnarray}
A steady-state power-law solution is sought in both the momentum and the multipole index. Accordingly, the solution is written as
\begin{eqnarray}
    n_{k,\ell m}=   Ak^{\nu}\ell^{\mu}.
\end{eqnarray}
It follows
\begin{eqnarray}
    &&\left(\frac{1}{n_{k,\ell m}}+\frac{1}{n_{3,\ell_3 m_3}}-\frac{1}{n_{1,\ell_1 m_1}}-\frac{1}{n_{2,\ell_2 m_2}}\right)n_{1,\ell_1 m_1}n_{2,\ell_2 m_2}n_{3,\ell_3 m_3}n_{k,\ell m}\to\nonumber\\
    &&(R_{k_3})^{3\nu}(R_{\ell_3})^{3\mu}\left(\frac{1}{n_{k,\ell m}}+\frac{1}{n_{\tilde{3},\tilde{\ell}_3\tilde{m}_3}}-\frac{1}{n_{\tilde{1},\tilde{\ell}_1\tilde{m}_1}}-\frac{1}{n_{\tilde{2},\tilde{\ell}_2\tilde{m}_2}}\right)n_{\tilde{1},\tilde{\ell}_1\tilde{m}_1} n_{\tilde{2},\tilde{\ell}_2\tilde{m}_2}n_{\tilde{3},\tilde{\ell}_3\tilde{m}_3}n_{k,\ell m}.
\end{eqnarray}
We consider a generalized dispersion relation \( \omega \sim k^{\alpha} \), noting that gravitational interaction in flat background selects \( \alpha = 1 \). The delta function will transform as follows
\begin{eqnarray}
    \delta(\omega^{3k}_{12})\to (R_{k_3})^{-\alpha}\delta(\tilde{\omega}^{k3}_{12}).
\end{eqnarray}
In the working assumption $\ell\gg 1$ we can write
\begin{eqnarray}
\sum_{\substack{\ell_1,\ell_2,\ell_3}} \to \prod_{i=1}^3\int d\ell_i.
\end{eqnarray}
In this case we can write again Eq. (\ref{secondint}) as
\begin{eqnarray}\label{secondint1}
\frac{1}{4}\prod_{i=1}^3\int d\ell_i\int dk_i k_i^2  
\Jijab{1}{2}{3}{k}n_{k,\ell m}n_{1,\ell_1 m_1}n_{2,\ell_2 m_2}n_{3,\ell_3 m_3}. 
\end{eqnarray}
This implies  a new Jacobian  from the multipoles integral
\begin{eqnarray}
d\ell_1d\ell_2d\ell_3\to(R_{\ell_3})^{4}d\tilde{\ell}_1d\tilde{\ell}_2d\tilde{\ell}_3.
\end{eqnarray}
Now we should transform the amplitude coefficient. Here we need to transform each term of Eq. (\ref{Aterm}). So we have
\begin{eqnarray}
W_{\substack{k_1,\ell'_1,m'_1\\k_2,\ell'_2,m'_2}}^{\substack{k_3,\ell',m'\\k_,\ell'_3,m'_3}}&=&\int d\Omega_k\int d\Omega_{1}\int d\Omega_{2}\int d\Omega_{3}\,W^{\mathbf{k_1},\mathbf{k_2}}_{\mathbf{k},\mathbf{k_3}}\,Y^*_{\ell' m'}(\Omega_k)Y^*_{\ell'_3 m'_3}(\Omega_3)Y_{\ell'_1 m'_1}(\Omega_1)Y_{\ell'_2 m'_2}(\Omega_2).
\label{wl}
\end{eqnarray}
Now applying the coordinate transformation (\ref{coordtransf}) we get 
\begin{eqnarray}
W_{\substack{R_{k_3}\tilde{k}_1,\ell'_1,m'_1\\R_{k_3}\tilde{k}_2,\ell'_2,m'_2}}^{\substack{R_{k_3}k,\ell',m'\\R_{k_3}\tilde{k}_3,\ell'_3,m'_3}}&=&\int d\Omega_k\int d\Omega_{1}\int d\Omega_{2}\int d\Omega_{3}\,W^{R_{k_3}\mathbf{\tilde{k}_1},R_{k_3}\mathbf{\tilde{k}_2}}_{R_{k_3}\mathbf{\tilde{k}_3},R_{k_3}\mathbf{k}}\,Y^*_{\ell' m'}(\Omega_k)Y^*_{\ell'_3 m'_3}(\Omega_3)Y_{\ell'_1 m'_1}(\Omega_1)Y_{\ell'_2 m'_2}(\Omega_2)\nonumber\\
&=&R_{k_3}^{\beta}\int d\Omega_k\int d\Omega_{1}\int d\Omega_{2}\int d\Omega_{3}\,W^{\mathbf{\tilde{k}_1},\mathbf{\tilde{k}_2}}_{\mathbf{\tilde{k}_3},\mathbf{k}}\,Y^*_{\ell' m'}(\Omega_k)Y^*_{\ell'_3 m'_3}(\Omega_3)Y_{\ell'_1 m'_1}(\Omega_1)Y_{\ell'_2 m'_2}(\Omega_2)\nonumber\\
&=&R_{k_3}^{\beta}W_{\substack{\tilde{k}_1,\ell'_1,m'_1\\\tilde{k}_2,\ell'_2,m'_2}}^{\substack{k,\ell',m'\\\tilde{k}_3,\ell'_3,m'_3}},
\label{wl1}
\end{eqnarray}
where we have assumed homogeneity of the scattering amplitude.
Next, we transform the delta function
\begin{eqnarray}
\delta_{\substack{k_1,\ell_1-\ell'_1,m_1-m'_1\\k_2,\ell_2-\ell'_2,m_2-m'_2}}^{\substack{k_3,\ell_3-\ell'_3,m_3-m'_3\\ k,\ell-\ell',m-m'}} &=& \int d\Omega_k \int d\Omega_1\int d\Omega_2\int d\Omega_3\; \delta(\mathbf{k}+\mathbf{k_3}-\mathbf{k_1}-\mathbf{k_2})\nonumber\\
 &\cdot&Y^*_{\ell-\ell', m-m'}(\Omega_k)Y^*_{\ell_3-\ell'_3, m_3-m_3'}(\Omega_3)Y_{\ell_1-\ell'_1, m_1-m_1'}(\Omega_1)Y_{\ell_2-\ell'_2, m_2-m_2'}(\Omega_2).
\end{eqnarray}
Considering the case in which $\ell_i\gg \ell'_i$ and $m_i\gg m'_i$ we get 

\begin{eqnarray}
 \delta_{\substack{R_{k_3}\tilde{k}_1,R_{\ell_3}\tilde{\ell}_1,R_{m_3}\tilde{m}_1\\R_{k_3}\tilde{k}_2,R_{\ell_3}\tilde{\ell}_2,R_{m_3}\tilde{m}_2}}^{\substack{R_{k_3}k,R_{\ell_3}\ell,R_{m_3}m\\ R_{k_3}\tilde{k}_3,R_{\ell_3}\tilde{\ell}_3,R_{m_3}\tilde{m}_3}} 
&\simeq& \int d\Omega_k \int d\Omega_1\int d\Omega_2\int d\Omega_3\; \delta(\mathbf{k}+\mathbf{k_3}-\mathbf{k_1}-\mathbf{k_2})\nonumber\\
 &\cdot&Y^*_{R_{\ell_3}\ell, R_{m_3}m}(\Omega_k)Y^*_{R_{\ell_3}\tilde{\ell}_3, R_{m_3}\tilde{m}_3}(\Omega_3)Y_{R_{\ell_3}\tilde{\ell}_1, R_{m_3}\tilde{m}_1}(\Omega_1)Y_{R_{\ell_3}\tilde{\ell}_2, R_{m_3}\tilde{m}_2}(\Omega_2).\nonumber\\
 &&
\end{eqnarray}
We can now expand the spherical harmonics for $\ell\gg1$ and fixed $m$. We recall the definition
\begin{eqnarray}
    Y_{\ell m}(\theta,\phi)=\sqrt{\frac{(2\ell+1)}{4\pi}\frac{(\ell-m)!}{(\ell+m)!}}P^m_{\ell}(\cos\theta)e^{im\varphi}.
\end{eqnarray}
In this limit, at first order in ${\cal O}(\ell^{-1})$, the Legendre polynomial $P^m_{\ell}(\cos\theta)$, behaves as
\begin{eqnarray}
P_\ell^{m}(\cos\theta) &=& \frac{(\ell+m)!}{(\ell-m)!}\ell^{-m} 
\left( \frac{\theta}{\sin\theta} \right)^{1/2} 
\left[ 
J_m\!\left(\left( \ell + \frac{1}{2} \right)\theta\right) \right].
\end{eqnarray}
It is mandatory to also expand the Bessel function $J_{m}\left[\left(\ell+\frac{1}{2}\right)\theta\right]$ 
\begin{equation}
J_m\!\left[\left(\ell + \tfrac{1}{2}\right)\theta\right] \sim 
\sqrt{\frac{2}{\pi\!\left(\ell + \tfrac{1}{2}\right)\theta}}\,
\cos\!\left[\left(\ell + \tfrac{1}{2}\right)\theta - \frac{m\pi}{2} - \frac{\pi}{4}\right].
\end{equation}
Since
\begin{eqnarray}
    \frac{(\ell+m)!}{(\ell-m)!}\simeq \ell^{2m},
\end{eqnarray}
at the lowest order in $\ell$, it is straightforward to obtain
\begin{eqnarray}
    Y_{\ell m}(\theta,\phi)\simeq \cos{\left[\left(\ell + \frac{1}{2}\right)\theta-\frac{m\pi}{2}-\frac{\pi}{4}\right]}e^{im\phi}.
\end{eqnarray}
Therefore we can transform the spherical harmonics
\begin{eqnarray}
    Y_{R_{\ell_3}\ell R_{m_3}m}(\theta,\phi)\simeq \cos\left[\left(R_{\ell_3}\ell + \frac{1}{2}\right)\theta-\frac{m\pi}{2}-\frac{\pi}{4}\right]e^{im\phi}\simeq  Y_{\ell m}(\theta,\phi).
\end{eqnarray}
Spherical harmonics are invariant under ZT transformations.
Using these asymptotic limits, we get 
\begin{eqnarray}
 \delta_{\substack{k_1,\ell_1-\ell'_1,m_1-m'_1\\k_2,\ell_2-\ell'_2,m_2-m'_2}}^{\substack{k_3,\ell_3-\ell'_3,m_3-m'_3\\ k,\ell-\ell',m-m'}}\simeq
 (R_{k_3})^{-3} \delta_{\substack{\tilde{k}_1,\tilde{\ell}_1-\ell'_1,\tilde{m}_1-m'_1\\\tilde{k}_2,\tilde{\ell}_2-\ell'_2,\tilde{m}_2-m'_2}}^{\substack{\tilde{k},\tilde{\ell}-\ell',\tilde{m}-m'\\\tilde{k}_3,\tilde{\ell}_3-\ell'_3,\tilde{m}_3-m'_3}}\,.
\end{eqnarray}
Now, as a last step, we need to transform the Clebsch-Gordan coefficients in the harmonic convolution.\\
We remember that
\begin{eqnarray}
    \mathcal{C}_{\ell,\ell_1,\ell_2,m,m_1,m_2} =\int d\Omega\, Y_{\ell m}Y_{\ell_1 m_1}Y_{\ell_2 m_2}
    = \sqrt{\frac{(2\ell+1)(2\ell_1+1)(2\ell_2+1)}{4\pi}}\left(\begin{matrix}
        \ell&\ell_1&\ell_2\\
        m&m_1&m_2
    \end{matrix}\right)\left(\begin{matrix}
        \ell&\ell_1&\ell_2\\
        0&0&0
    \end{matrix}\right),
\end{eqnarray}
where $\left(\begin{matrix}a&b&c\\d&e&f
\end{matrix}\right)$ is the 3j Wigner symbol. In the limit for $\ell\gg1$ we can use the Ponzano-Regge relation \cite{Ponzano1969SEMICLASSICALLO}, getting
\begin{eqnarray}
    \left(\begin{matrix}
        \ell&\ell_1&\ell_2\\
        m&m_1&m_2
    \end{matrix}\right) \simeq \sqrt{\frac{1}{12\pi V}}\cos{\left[\sum_i \left(\ell_i+\frac{1}{2}\right)\theta + \frac{\pi}{4}\right]},
\end{eqnarray}
where $V$ is the volume of the tetrahedra in the harmonic space, while 
\begin{eqnarray}
    \left(\begin{matrix}
        \ell&\ell_1&\ell_2\\
        0&0&0
    \end{matrix}\right)\simeq \rm{const.}\,,
\end{eqnarray}
in the same limit.
Therefore, we can easily see that the Clebsch-Gordan coefficient remains invariant for the coordinate transformation performed . Another way to spot the invariance of the Clebsch-Gordan coefficient is by noticing that spherical harmonics are invariant under ZT transformations in the eikonal limit. At the end we can write the total transformation for the whole amplitude factor as
\begin{eqnarray}\label{Achange}
    \left|
    \LA \right|^2\to (R_{k_3})^{2\beta-3}\left|
    \Aijab{\tilde 1}{\tilde 2}{\tilde k}{\tilde 3}\right|^2.
\end{eqnarray}
Combining all together, we have
\begin{eqnarray}
&&\frac{1}{4}\prod_{i=1}^3\int d\ell_i\int dk_i k_i^2  
\Jijab{1}{2}{3}{k}n_{k,\ell m}n_{1,\ell_1 m_1}n_{2,\ell_2 m_2}n_{3,\ell_3 m_3}=\nonumber\\
&&\frac{1}{4}\prod_{i=1}^3\int d\tilde\ell_i\int d\tilde{k}_i \tilde{k}_i^2  
\Jijab{\tilde 1}{\tilde 2}{k}{\tilde 3}n_{k,\ell m}n_{\tilde{1},\tilde{\ell}_1 \tilde{m}_1}n_{\tilde{2},\tilde{\ell}_2 \tilde{m}_2}n_{\tilde{3},\tilde{\ell}_3 \tilde{m}_3}\left(\frac{\tilde{k}_3}{k}\right)^x\left(\frac{\tilde{\ell_3}}{\ell}\right)^y
\end{eqnarray}
with 
\begin{eqnarray}\label{xy}
    x&=&\alpha-3\nu-2\beta-9,\nonumber\\
    y&=&-3\mu-4.
\end{eqnarray}
Now we have to perform a specific ZT transformation for the third and fourth integral of Eq. (\ref{boltzsymm}). In this way we get
\begin{eqnarray}
\dot{n}_{k,\ell m} &=& \frac{1}{4}\prod_{i=1}^3\int d\ell_i\int dk_i k_i^2  n_{k,\ell m}n_{1,\ell_1 m_1}n_{2,\ell_2 m_2}n_{3,\ell_3 m_3}\,
\Jijab{1}{2}{k}{3}\nonumber\\ &\cdot&\left[1+\left(\frac{k_3}{k}\right)^x\left(\frac{\ell_3}{\ell}\right)^y-\left(\frac{k_1}{k}\right)^x\left(\frac{\ell_1}{\ell}\right)^y-\left(\frac{k_2}{k}\right)^x\left(\frac{\ell_2}{\ell}\right)^y \right].
\label{kinl}
\end{eqnarray}
\vskip 0.5cm

\section{ The nonlinear time scale}
\noindent
One point worth checking is the time scale it takes for the interactions to occur as the system should have time to develop a turbulent behavior. Notice first that we are dealing with gravitational strains and not with dimensionful graviton fields. For the simplicity's sake, let us estimate the interaction time in the case of a trilinear graviton strain coupling $\sim h(\partial h)^2$ (to be consistent with the main text, one should estimate the nonlinear time for the quartic coupling, but we expect a similar estimate as the quartic interactions goes like $h^2(\partial h)^2$). Furthermore, let us focus on a system where large multipole perturbations have been injected and calculate what time it takes to have a cascade generating a strain $h_{\rm low}$ with much smaller multipole. The calculation is similar to what is done to compute the nonlinear gravitational wave memory effect. 

The low multipole mode 
solves the back-reaction equations \cite{Maggiore:2007ulw}
\begin{eqnarray}\label{eq:EinEqmem}
    G_{\mu\nu}[h_{\rm low}]= 8\pi T^{\rm gw}_{\mu\nu}, 
\end{eqnarray}
where the sourced term on the right-hand side is the pseudo
energy-momentum tensor of the primary high multipole GWs 

\begin{eqnarray}\label{eq:TGWs}
    T^{\rm gw}_{\mu\nu}= 
    \frac{n_\mu n_\nu}{r^2}\frac{\mathrm{d}E(u,\Omega)}{\mathrm{d}u\mathrm{d}\Omega}=
    \frac{n_\mu n_\nu}{16\pi}\langle \dot{h}_+^2+\dot{h}_\times^2\rangle. 
\end{eqnarray}
 By defining, as usual, $\Bar{h}_{\mu\nu}=h_{\mu\nu}-\eta_{\mu\nu}h^\alpha_\alpha/2$, in the harmonic gauge where $\partial^\mu \bar{h}_{\mu\nu}=0$ for the low frequency part as well, Eq. \eqref{eq:EinEqmem} becomes
\begin{eqnarray}
    \Box \Bar{ h}^{\rm low}_{\mu\nu}=-16\pi T^{\rm gw}_{\mu\nu} .
\end{eqnarray}
This can be solved by the usual Green function method such that at position $\Vec{x}$ it is given by the integral over the whole space where the source is non-zero
\begin{eqnarray}\label{eq:memsteps}
    \Bar{ h}^{\rm low}_{\mu\nu}(t,\vec{x})&=& \int {\rm d}r'{\rm d}\Omega' {\rm d}u' n'_\mu n'_\nu\frac{\mathrm{d}E(u',\Omega')}{\mathrm{d}u'\mathrm{d}\Omega'}\nonumber\\
    &\cdot&\frac{\delta(t-t'-r|1-\Hat{n}\cdot \Vec{r'}/r|)}{r|1-\Hat{n}\cdot \Vec{r'}/r|}  .
\end{eqnarray} 
We expand
both the  contribution into spin-weighted spherical harmonics  of spin-weight $s=-2$, with $u=(t-r)$,  
\begin{eqnarray}\label{eq:SW Memory quantity GR} 
     h(u,r,\Omega)\equiv  h_+-ih_\times=\sum_{\ell=2}^\infty\sum_{m=-\ell}^{\ell}\, h_{\ell m}(u,r)\,\,_{{-2}}Y_{\ell m}(\Omega),
\end{eqnarray}
where the spatial direction is parametrized by the two angles in the reference frame centered at the source.
 Taking this decomposition the energy flux can be written as  
\begin{eqnarray}
    |\dot h|^2=\sum_{\ell_1=2}^\infty\sum_{m_1=-\ell_1}^{\ell_1}\sum_{\ell_2=2}^\infty\sum_{m_2=-\ell_2}^{\ell_2} \dot h_{\ell_1m_1}\dot h^*_{\ell_2m_2}\,_{{-2}}Y_{\ell_1m_1}\,_{{-2}}Y^*_{\ell_2m_2}\,,
\end{eqnarray}
one can compute the low frequency GW  from the modes of the primary wave. 
The TT projection factor can be also written into spherical harmonics \cite{Favata:2008yd} 
\begin{eqnarray}
    h_{\ell m}^{\rm low}=r\sum_{\ell',\ell''\geq 2}\, \sum_{m',m''} \Gamma^{\ell'm'm''\ell''}_{\ell m} \int_{-\infty}^u \mathrm{d}u'\langle\dot{h}^{\ell' m'} \dot{h}^{*\ell'' m''}\rangle,
\label{eq:memorymodes}
\end{eqnarray}
with   
\begin{eqnarray}
    \Gamma^{\ell'm'm''\ell''}_{\ell m}&\equiv& (-1)^{m+m''}\sqrt{\frac{(2\ell'+1)(2\ell''+1)(2\ell+1)}{4\pi}}\nonumber\\
    &\times&\sqrt{\frac{(\ell-2)!}{(\ell+2)!}}
    \begin{pmatrix}
        \ell' & \ell'' & \ell\\
         m' & -m'' & -m
    \end{pmatrix}
    \begin{pmatrix}
        \ell' & \ell'' & \ell\\
        2 & -2 & 0
    \end{pmatrix},
\end{eqnarray}
where the big parenthesis represents the Wigner $3j$ symbols, which in this case are only non-zero if $m=m'-m''$ and $|\ell'-\ell''|\leq \ell\leq \ell'+\ell''$. If we take $\ell=m=2$, $\ell'=m'$ and $\ell''=m''=\ell'+2$, we obtain the coefficient to be 0.78 for $\ell',\ell''\gg 2$. In the same limit, for the fundamental mode,  we have (we set to unity Newton's constant)

\begin{eqnarray}
r h_\ell=A_\ell e^{-i\omega_\ell u},\quad\omega_\ell=\omega\left[\left(\ell+\frac{1}{2}\right)-\frac{i}{2}\right],\quad\omega=\frac{1}{3\sqrt{3} M},
\end{eqnarray}
and

\begin{eqnarray}
r h_{22}^{\rm low}\simeq -0.78 A_\ell A_{\ell+2}\omega_\ell\omega_{\ell+2}\frac{\left(1+2i\right)}{5\omega}e^{-2i\omega u-\omega u}.
\end{eqnarray}
The time-scale $\tau_{\rm NL}$ to initiate the cascade is therefore of the order of $ h_{22}^{\rm low}/ \dot{h}_{22}^{\rm low}\sim (1/2\omega)\sim M$  and it will last till the initial large multipole driver  dies off, i.e. till $(1/{\rm Im}\,\omega_\ell)$. On the other hand $\tau_{\rm L}\sim 1/\omega_\ell\sim (M/\ell)$, so that $\epsilon^4\sim 1/\ell$.

\end{document}